\documentclass{ar2e}
\usepackage{amssymb}
\usepackage{psfig}

\newcommand{\prd}{\emph{Phys. Rev. D.}}
\newcommand{\apj}{\emph{Ap. J.}}

\def\n{{\bf n}}

\def\muK{{\rm \mu K}}

\def\bi{\begin{itemize}}
\def\ei{\end{itemize}}

\def\n2{N$_{2}$}
\def\n2{N$_{2}$}
\def\4he{$^{4}$He}
\def\4he{$^{4}$He}
\def\cm3{cm$^3$}

\def\cm3{cm$^3$}

\def\lya{Ly$\alpha$ }
\def\lyb{Ly$\beta$ }
\def\lyc{Ly$\gamma$ }
\def\strom{Str\"{o}mgren~}
\def\sfrd{\,{\rm M_\odot\,yr^{-1}\,Mpc^{-3}}}
\def\aj{A.J.}
\def\apj{Ap.J.}
\def\mnras{MNRAS}
\def\nat{Nature}
\def\apjl{Ap.J. Lett.}
\def\mnras{MNRAS}
\def\aap{A\&A}
\def\pasj{PASJ}
\def\araa{ARA\&A}
\def\procspie{SPIE}

\newcommand{\arcdeg}{^\circ}

\usepackage{ulem}  

\begin{document}

\input epsf.tex    
\input epsf.def   

\input psfig.sty

\jname{Annu. Rev. Astro. Astrophy.}
\jyear{2005}
\jvol{1}
\ARinfo{1056-8700/97/0610-00}

\title{Observational constraints on Cosmic Reionization}

\markboth{Fan, Carilli, Keating}{Cosmic Reionization}

\author{Xiaohui Fan$^1$, C.L. Carilli$^2$, B. Keating$^3$
\affiliation{$^1$Steward Observatory, University of Arizona, Tucson,
AZ 85721; \\
$^2$National Radio Astronomy Observatory, Socorro, NM, 87801; \\
$^3$Department of Physics, University of California, San Diego,
CA, 92093}}

\begin{keywords}
{Cosmology -- observational; Galaxy formation; cosmic reionization}
\end{keywords}

\begin{abstract}

Recent observations have set the first constraints on the epoch of
reionization (EoR), corresponding to the formation epoch of the first
luminous objects. Studies of Gunn-Peterson (GP) absorption, and
related phenomena, suggest a qualitative change in the state of the
intergalactic medium (IGM) at $z \sim 6$, indicating a rapid increase
in the neutral fraction of the IGM, from $x_{HI} < 10^{-4}$ at $z \le
5.5$, to $x_{HI} > 10^{-3}$, perhaps up to 0.1, at $z \ge 6$.
Conversely, transmission spikes in the GP trough, and the evolution of
the \lya galaxy luminosity function indicate $x_{HI} < 0.5$ at $z\sim
6.5$, while the large scale polarization of the cosmic microwave
background (CMB) implies a significant ionization fraction extending
to higher redshifts, $z \sim 11 \pm 3$. The results suggest that
reionization is less an event than a process, with the process
beginning as early as $z \sim 14$, and with the 'percolation', or
'overlap' phase ending at $z \sim 6$.  The data are consistent with
low luminosity star forming galaxies as being the dominant sources of
reionizing photons.  Low frequency radio telescopes currently under
construction should be able to make the first direct measurements of
HI 21cm emission from the neutral IGM during the EoR, and upcoming
measurements of secondary CMB temperature anisotropy will provide fine
details of the dynamics of the reionized IGM.

\end{abstract}

\maketitle

\section{Introduction}

The baryonic pre-galactic medium (PGM) evolves in three distinct
phases. At high redshifts ($z > 1100$) the PGM is hot, fully ionized,
and optically thick to Thomson scattering, and hence coupled to the
photon field. As the universe expands, the PGM cools, and eventually
recombines, leaving a surface of last scattering (the cosmic microwave
background, CMB), plus a neutral PGM.  This neutral phase lasts from
$z = 1100$ to $z \sim 14$.  At some point between $z \sim 14$ and 6,
hydrogen in the PGM is `reionized', due to UV radiation from the first
luminous objects, leaving the fully reionized intergalactic medium
(IGM) seen during the `realm of the galaxies' ($6 > z > 0$). The
ionized, dense PGM at very high redshift has been well studied through
extensive observations of the CMB. Likewise, the reionized, rarified
IGM at low redshift has been well characterized through QSO absorption
line studies. The middle phase -- the existence of a neutral IGM
during the so-called `dark ages' (Rees 1998), and the process of
reionization of this medium, is the last directly observable phase of
cosmic evolution that remains to be verified and explored.  The epoch
of reionization (EoR) is crucial in cosmic structure formation
studies, since it sets a fundamental benchmark indicating the
formation of the first luminous objects, either star forming galaxies
or active galactic nuclei (AGN).

Cosmic reionization has been discussed within the larger context of
mostly theoretical reviews on the formation of the first luminous
objects (Loeb \& Barkana 2002; Barkana \& Loeb 2001; Ciardi \& Ferrara
2005).  The past few years have seen the first observational evidence
for a neutral IGM, and the first constraints on the process of
reionization.  In this review we do not repeat the theory of early
structure formation, but focus on these recent observations of
reionization and the evolution of the neutral IGM.  Such constraints
come from optical and near-IR spectroscopy of the most distant QSOs,
and study of galaxy populations at the highest redshifts, among
others. Detections of the IGM Thomson optical depth come from
observations of the large angular scale polarization of the CMB (Kogut
et al. 2003; Spergel et al. 2006; Page et al. 2006").
These measurements yield a $\gtrsim 3\sigma$
detection of polarized emission of the CMB,
and future observations of the small scale CMB anisotropy will
complement the large angular scale polarization by probing the
fine-details of the reionization epoch. We also review the potential
sources of reionization, including star forming galaxies, AGN, and
decaying particles. We conclude with a discussion of low frequency
radio telescopes that are being constructed to detect the neutral IGM
directly through the 21cm line of neutral hydrogen.

We focus our review on the observations of hydrogen
reionization. Neutral helium reionization likely happened at a similar
epoch to that of neutral hydrogen, due to their similar ionization
potential and recombination rate.  Reionization of He II happens at a
much lower redshift -- the He II Gunn-Peterson effect is observed in
the spectrum of quasars at $z\sim 3$ (e.g. Jakobsen et al. 1994,
Davidsen, Anderson et al. 1999, Kriss et
al. 2001, Shull et al. 2004, Zheng et al. 2004).  Evolution of the IGM
temperture (e.g. Ricotti, Gnedin \& Shull 2000, Schaye et al. 2000,
Theuns et al. 2002a,b) also suggest a rapid change of IGM temperture
at $z\sim 3$, consistent with the onset of He II reionization.  For a
detailed recent review of HeII reionization, see Ciardi \& Ferrara
(2005).

\section{A basic model of reionization}

In order to put recent observations in context, we include a brief,
mostly qualitative, description of the reionization process.  Again,
for more extensive reviews of early structure formation, see Loeb \&
Barkana (2002) Barkana \& Loeb( 2002), and Ciardi \& Ferrara (2005).

Assuming reionization is driven by UV photons from the first luminous
sources (stars or AGN), analytic (eg. Madau, Haardt, \& Rees 1997;
Miralda-Escude, Haenelt, \& Rees 2000; Wyithe \& Loeb 2003), and
numerical (eg. Gnedin 2000; Razoumov et al. 2002; Sokasian et
al. 2003; Ciardi, Stoehr, \& White 2004; Paschos \& Norman 2005),
calculations suggest that reionization follows a number of phases. The
first phase will be relatively slow, with each UV source isolated to
its own \strom   sphere. Eventually, these spheres grow and join,
leading to much faster reionization, due to the combination of
accelerating galaxy formation, plus the much larger mean free path
of ionizing photons in the now porous IGM. This stage is known as
`overlap', or `percolation'. The last phase entails the etching-away
of the final dense filaments in the IGM by the intergalactic UV
radiation field, leading to a fully ionized IGM (neutral fraction,
$x_{HI} \def \frac{n_{\rm HI}}{n_{\rm H}}
\sim 10^{-5}$, at $z = 0$).

There are a number of important caveats to this most basic model.
First, causality, cosmic variance, and source clustering will set an
absolute minimum width in redshift for the fast phase of reionization,
$\Delta z > 0.15$, and the process itself is likely to be much more
extended in time (Wyithe \& Loeb 2004). In particular, Furlanetto \&
Oh (2005) point out that the overlap phase is a relatively local
phenomenon, with significant differences along different lines of
sight.  Second, the recombination time for the mean IGM becomes longer
than the Hubble time at $z \ge 8$, allowing for the possibility of
complex reionization histories.  As shown by a number of
semi-analytical works (e.g., Cen 2003a,b, Chiu et al. 2003, Wyithe \&
Loeb 2003, Haiman \& Holder 2003), adding feedback from an early
generation of galaxy formation, or adjusting star formation efficiency
at $z>6 - 10$, will produce a prolonged, or multi-epoch reionization
history.  Third, it remains an open question as to whether
reionization proceeds from high to low density regions (inside-out;
Ciardi \& Madau 2003, Sokasian et al. 2003; Iliev et al. 2006), or
from low to high density regions (outside-in; Miralda-Escude et
al. 2000, Gnedin 2000). Forth, a very different picture arises if we
assume reionization by decaying fundamental particles, or by
penetrating hard Xrays from the first luminous sources.  In this case,
the reionization process is not localized to individual \strom
spheres, but occurs more uniformly throughout the IGM (\S 8.4).
Lastly, we have been necessarily vague in terms of precise redshifts
and timescales for the various phases -- this is the main concern of
the present review.

Throughout this review, we adopt a standard concordance cosmology
(Spergel et al. 2003, Spergel et al. 2006), unless stated otherwise.

\section{Observations of the Gunn-Peterson Effect at $z\sim 6$}

Gunn \& Peterson (1965) first proposed using \lya resonance absorption
in the spectrum of distant quasars as a direct probe to the neutral
hydrogen density in the IGM at high-redshift (see also Field 1959, Shklovsky 1964,
Scheuer 1965, Bahcall \& Salpeter 1965).  For objects beyond
reionization, neutral hydrogen in the IGM creates complete
Gunn-Peterson absorption troughs in the quasar spectrum blueward of
Ly$\alpha$ emission. Observations of the Gunn-Peterson effect
directly constrain the evolution of neutral
hydrogen fraction and the ionization state of the IGM.

\subsection{Gunn-Peterson Effect: Basics}

The Gunn-Peterson (1965) optical depth to \lya photons is
\begin{equation}
\tau_{\rm GP} = \frac{\pi e^2}{m_e c} f_{\alpha} \lambda_{\alpha} H^{-1}(z) n_{\rm HI },
\end{equation}
where $f_{\alpha}$ is the oscillator strength of the \lya transition,
$\lambda_\alpha$ = 1216\AA, $H(z)$ is the Hubble constant at redshift
$z$, and $n_{\rm HI }$ is the density of neutral hydrogen in the IGM.
At high redshifts:
\begin{equation}
\tau_{\rm GP} (z) = 4.9 \times 10^5 \left( \frac{\Omega_m h^2}{0.13} \right)^{-1/2}
\left( \frac{\Omega_b h^2}{0.02} \right)
\left ( \frac{1+z}{7} \right )^{3/2}
\left( \frac{n_{\rm HI}}{n_{\rm H}} \right ),
\end{equation}
for a uniform IGM.
Even a tiny neutral fraction,
$x_{HI} \sim 10^{-4}$, gives rise to
complete Gunn-Peterson absorption.
This test is only sensitive at the end of the reionization
when the IGM is already mostly ionized, and saturates
for the higher neutral fraction in the earlier stage.

At $z<5$, the IGM absorption is resolved into individual \lya forest
lines; their number density increases strongly with redshift ($N(z)
\propto (1+z)^{~2.5}$, see Rauch 1998).  Earlier attempts to study
Gunn-Peterson absorption concentrated on measuring the amount of flux
between individual \lya forest lines using high-resolution
spectroscopy to place limits on the diffuse neutral IGM.
Webb et al. (1992) and Giallongo et al. (1994) found
$\tau < 0.05$ at $z\sim 4$.  Songaila et al. (1999) and Fan et al. (2000)
place upper limits of $\tau = 0.1$ and 0.4 based on lack of complete Gunn-Peterson
troughs in SDSS
J0338+0021 ($z=5.00$) and SDSS J1044--0125 ($z=5.74$), respectively.
However, at $z>5$, even with a
moderately high resolution spectrum, \lya forest lines overlap
severely, making it impossible to find a truly "line-free" region.

A more accurate picture of the IGM evolution interprets the \lya
forest as a fluctuating Gunn-Peterson effect: absorption arises from
low-density gas in the IGM that is in approximate thermal equilibrium
between photoionization heating by the UV background and adiabatic
cooling due to the Hubble expansion, rather than as discrete \lya
forest clouds (Bi 1993; Cen et al. 1994; Zhang et al. 1995; Herquist
et al. 1996).  The neutral hydrogen fraction and
therefore the GP optical depth depends on the local density of the
IGM.  By studying the evolution of the average transmitted flux or
effective optical depth, one can trace the evolution of the UV
ionizing background and neutral fraction of the IGM.  At
high-redshift, the IGM is highly clumpy (Miralda-Escude et al. 2000),
which has to be taken into account in order to estimate the IGM
ionization from observations.

\subsection{Complete Gunn-Peterson troughs: phase transition or gradual
evolution?}

\begin{figure}
\centerline{\psfig{figure=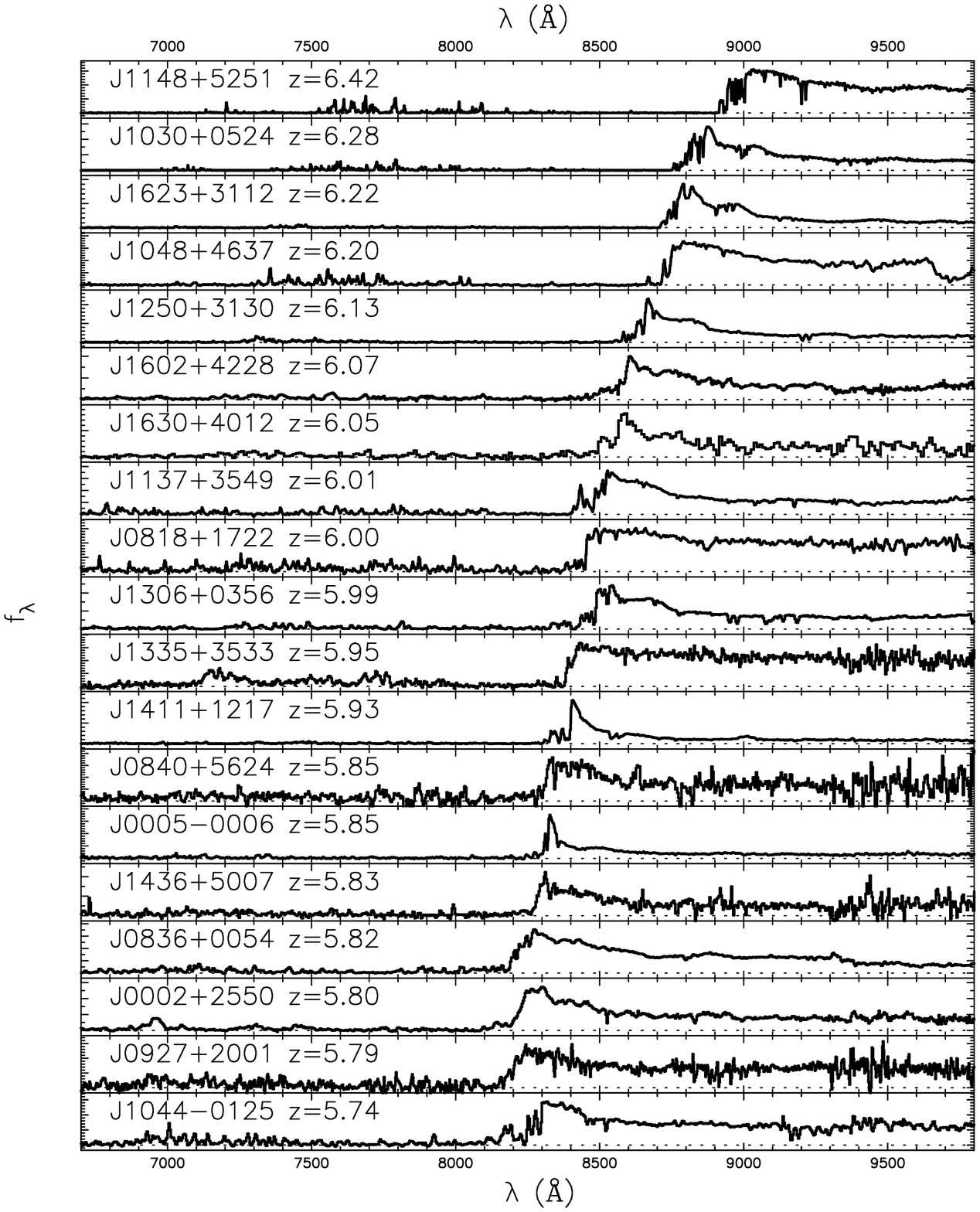,height=30pc}}
\caption{Moderate resolution spectra of nineteen SDSS quasars at $5.74 < z < 6.42$.
Adapted from Fan et al. (2006b)}
\label{z19}
\end{figure}

The Sloan Digital Sky Survey (SDSS, York et al. 2000)
provides large samples of luminous quasars over $0 < z <
6.5$.  Fan et al. (2000,
2001, 2003, 2004, 2006a) carried out a survey of $i$-dropout quasars
($z>5.7$) using the SDSS imaging data, resulting in the discovery of 19
luminous quasars in this redshift regime (Figure \ref{z19}).  Other
multicolor survey projects (e.g, Mahabal et al. 2005; AGES, Cool et al. 2006;
QUEST, Djorgovski et al. 2005; CFHT Legancy
survey, Willott et al. 2005) are also searching for
quasars at similar redshifts. They provide by far the
best probes of IGM evolution towards the end of the EoR.

\begin{figure}
\centerline{\psfig{figure=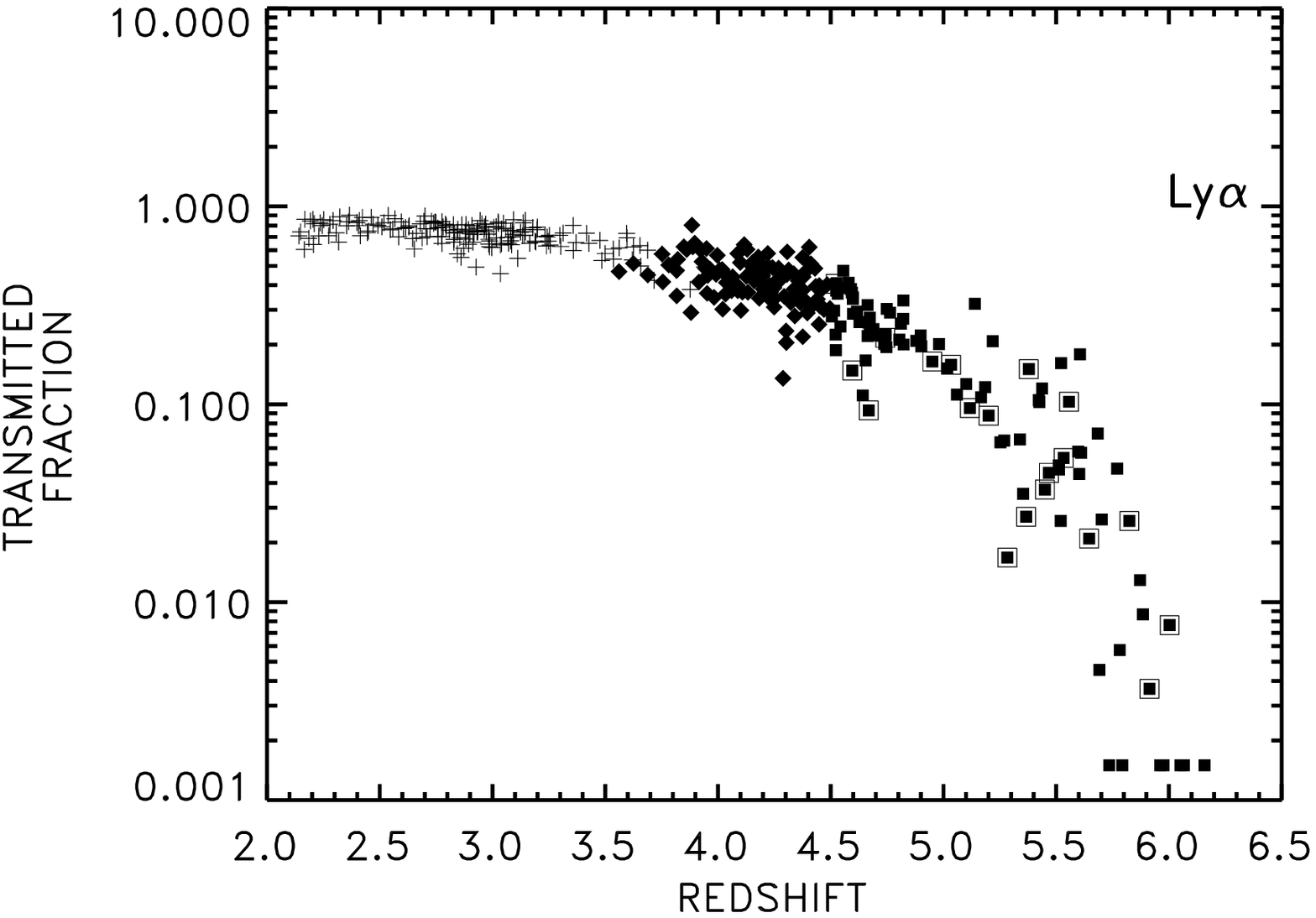,height=20pc}}
\caption{Transmitted flux blueward of \lya emission as a
function of redshift from $z\sim 2$ to 6.3 using
Keck/ESI and HIRES data. Large open squares are points
data for possible BAL quasars. Flux was computed in bins of 15\AA.
Adapted from Songaila (2004)}
\label{songaila1}
\end{figure}

Songaila (2004) summarized the evolution of transmitted flux over a
wide redshift range ($2 < z < 6.3$) using high signal-to-noise,
moderate-resolution ($R\ge 5000$) observations of 50 quasars.  Strong
evolution of \lya absorption at $z_{\rm abs} > 5$ is evident, the
transmitted flux quickly approaches zero at $z>5.5$ (Figure
\ref{songaila1}).
At $z_{\rm abs} > 6$,
complete absorption troughs begin to appear: the Gunn-Peterson
optical depths are $>>1$, indicating a rapid increase in the neutral
fraction.  However, is this evolution the result of a smooth transition
due to gradual thickening of the \lya forest (e.g. Songaila \& Cowie 2002,
Songaila 2004)? Or is it a reflection of a more dramatic change in the
IGM evolution, marking the end of reionization (e.g., Djorgovski et
al. 2001, Becker et al. 2001, Fan et al. 2002)?

\begin{figure}
\centerline{\psfig{figure=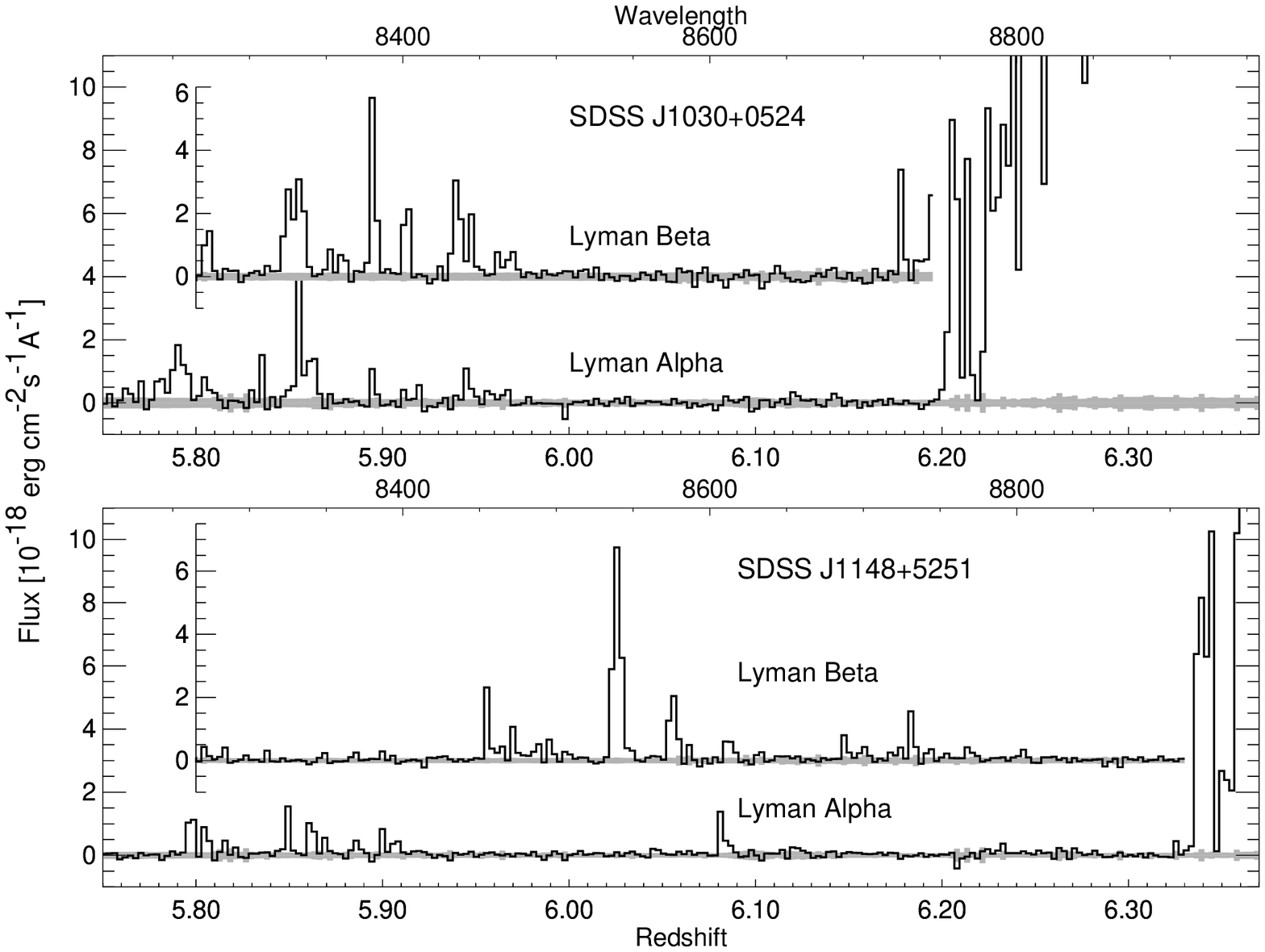,height=20pc}}
\caption{Close-up of the \lya and \lyb troughs in the two highest redshift
quasars currently known. SDSS J1030+0524 ($z=6.28$) shows complete Gunn-Peterson
absorption in both \lya and \lyb, where SDSS J1148+5251 ($z=6.42$) has
clear transmission, suggesting large line of sight variance at the end
of reionization. Adapted from White et al. (2003). }
\label{white1}
\end{figure}

Djorgovski et al. (2001) detected an extended dark gap in the spectrum
of SDSS J1044--0125 ($z=5.74$), at $z_{\rm abs} = 5.2 - 5.6$, with
$\tau_{\rm GP} > 4.6$.  The first clear-cut Gunn-Peterson trough was
discovered in the spectrum of SDSS J1030+0524 ($z=6.28$, Fan et
al. 2001, Becker et al. 2001, Figure \ref{white1}), which showed
complete Gunn-Peterson absorption at $5.95 < z_{\rm abs} < 6.15$ in
both \lya and Ly$\beta$ transitions.  A high S/N spectrum presented in
White et al. (2003) placed a stringent limit $\tau_{GP} > 6.3$ in
\lya.  However, Songaila (2004) suggested that constraints based on
\lya alone do not deviate from a simple extrapolation from lower
redshift by a large factor, consistent with a relatively smooth
evolution in the ionization state.

The GP optical depth $\tau \propto f\lambda$, where $f$ and $\lambda$
are the oscillator strengh and rest-frame wavelength of the transition.
For the same neutral density, the GP optical depth of \lyb and \lyc are
factors of 6.2 and 17.9 smaller than that of Ly$\alpha$.
Therefore,  \lyb can provide more stringent constraints on
the IGM ionization state when \lya absorption saturates.
Complete \lyb absorption implies a
significant rise in the neutral fraction towards SDSS J1030+0524.
Songaila (2004) pointed out that in a clumpy IGM, the simple
conversion factor of 6.2 between \lya and \lyb optical depths is an
upper limit.
Using high order Lyman lines requires proper treatment of the clumpy
IGM. Oh \& Furlanetto (2005) and Fan et
al. (2006b) found $\tau^{\alpha} /
\tau^{\beta} \sim 2 - 3 $ and $\tau^{\alpha} / \tau^{\gamma} \sim 4 -
6 $ for a clumpy IGM.

\begin{figure}
\centerline{\psfig{figure=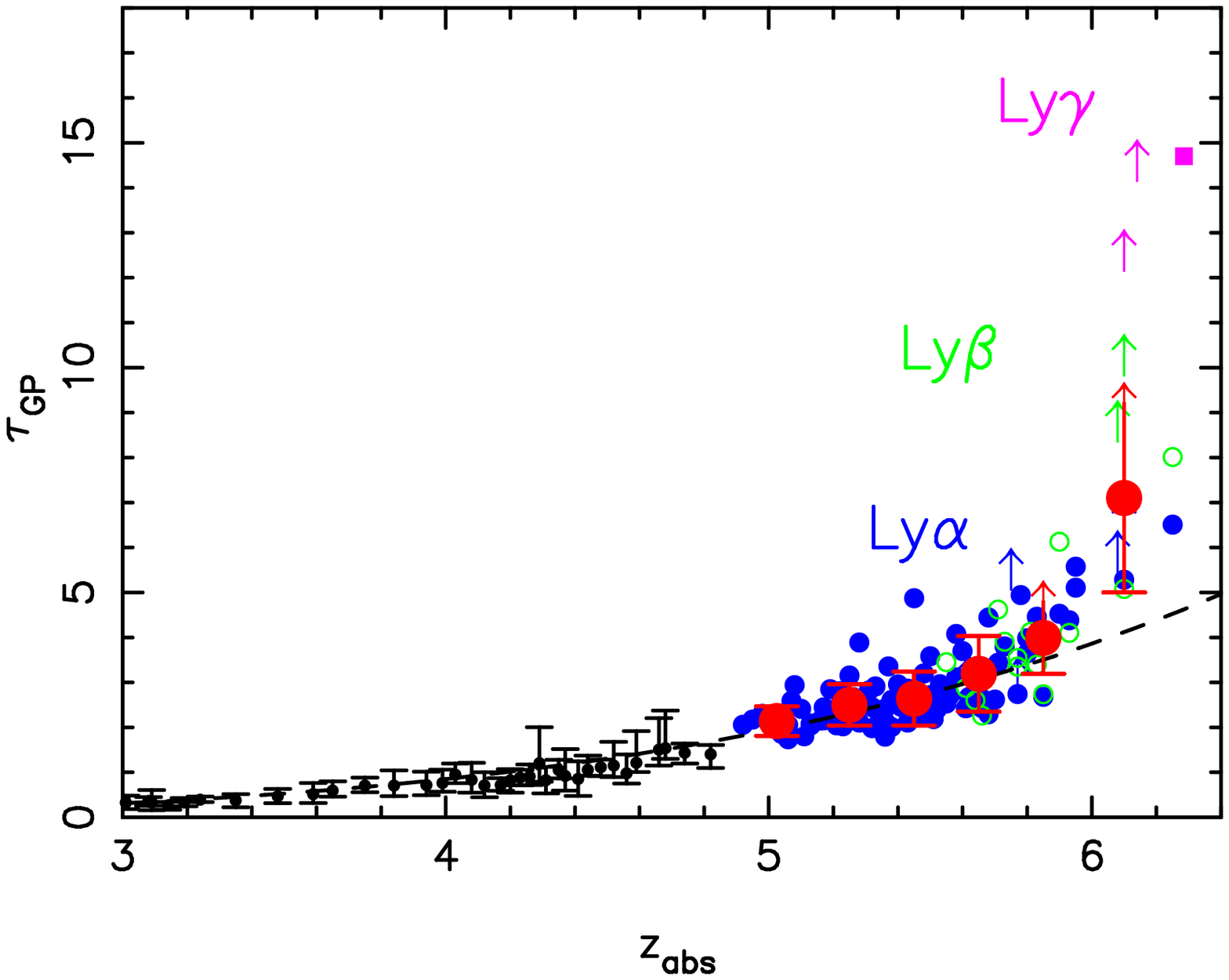,height=20pc}}
\caption{Evolution of Optical depth with combined \lya and \lyb results.
The dash line is for a redshift evolution of
$\tau_{\rm GP} \propto (1+z)^{4.3}$. At $z>5.5$,
the best fit evolution has
$\tau_{\rm GP} \propto (1+z)^{>10.9}$,
indicating an accelerated evolution.
The large open symbols with error bars
are the average and standard deviation
of optical depth at each redshift. The
sample variance increases also increases
rapidly with redshift. Adapted from Fan et al. (2006b).}
\label{alpha+beta}
\end{figure}

Fan et al. (2006b) measured the evolution of Gunn-Peterson optical
depths along the line of sight of the nineteen $z>5.7$ quasars from
the SDSS (Figure \ref{alpha+beta}).  They found that at $z_{\rm
abs}<5.5$, the optical depth can be best fit as $\tau \propto
(1+z)^{4.3}$, while at $z_{\rm abs}>5.5$, the evolution of optical
depth accelerates: $\tau \propto (1+z)^{>10}$.  There is also a rapid
increase in the variation of optical depth along different lines of
sight: $\sigma(\tau)/\tau$ increases from $\sim 15$\% at $z\sim 5$, to
$>30\%$ at $z>6$, in which $\tau$ is averaged over a scale of $\sim
60$ comoving Mpc.  Figure \ref{white1} (White et al. 2003) compares
the Gunn-Peterson absorption in the two highest redshift quasars at
the time of this writing: SDSS J1030+0524 ($z=6.28$) shows complete
absorption.  On the other hand, SDSS J1148+5251 ($z=6.42$) shows clear
transmission in \lya and \lyb transitions at the same redshift range,
clearly indicating that that this line of sight is still highly
ionized (Oh \& Furlanetto 2005, White et al. 2005).  The increased
variance in the IGM optical depth at the highest redshifts is further
discussed in Songaila (2004), Djorgovski et al. (2005) and Fan et
al. (2006b).

Wyithe \& Loeb (2004) and Furlanetto \& Oh (2005) pointed out that at
the end of reionization, the size of ionized bubbles grow
rapidly. They estimate an observed bubble size of tens of Mpc at
$z\sim 6$, and a scatter in the observed redshift of overlap along
different lines-of-sight to be $\Delta z \sim 0.15$.  Wyithe \& Loeb
(2005a) further showed that the Gunn-Peterson optical depth variations
to be of order unity on a scale of $\sim 100$ comoving Mpc. The
increased variance on the IGM optical depth supports the
interpretation that the IGM is close to the end of reionization.

\begin{figure}
\centerline{\psfig{figure=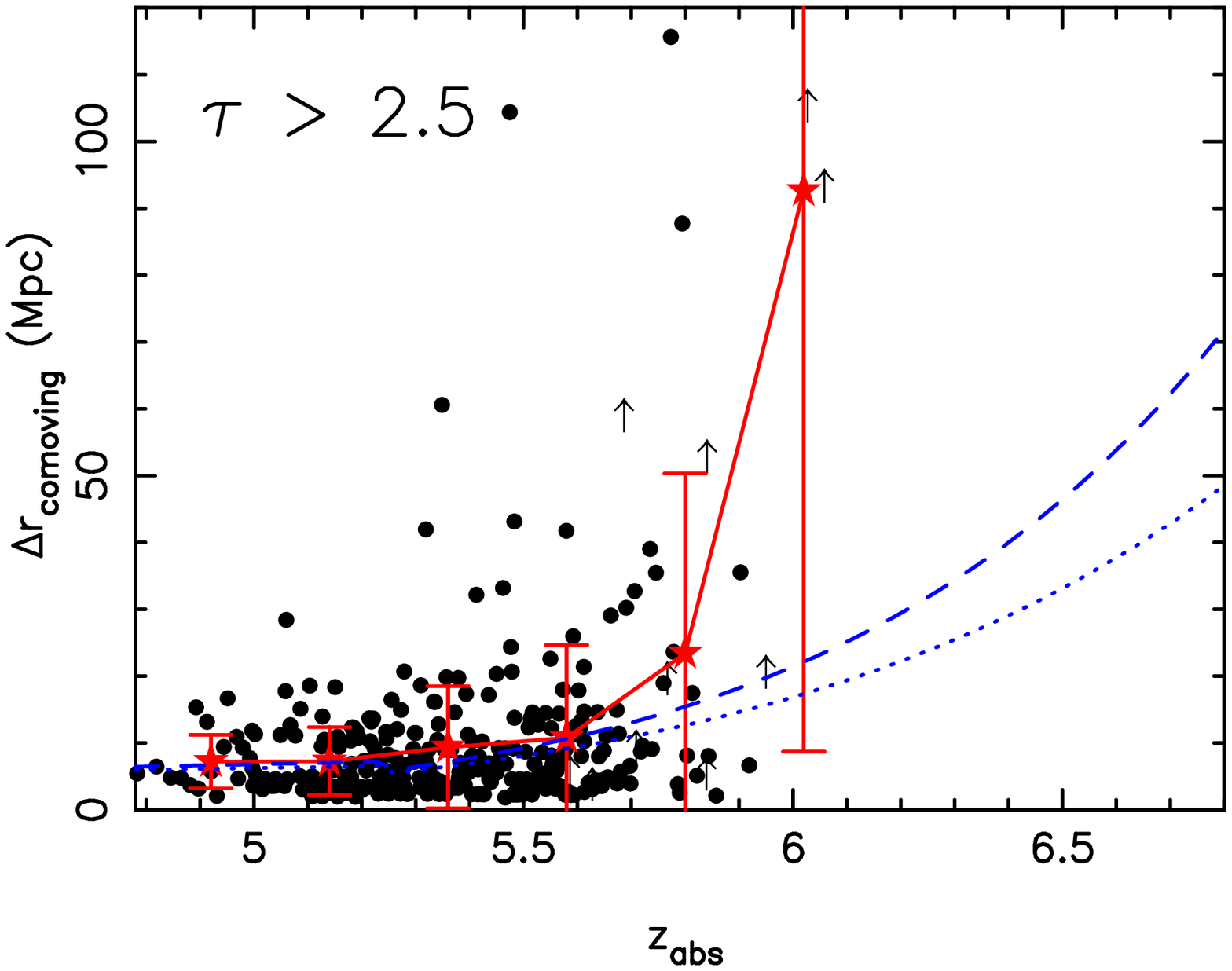,height=20pc}}
\caption{Distributions of dark gaps, defined as
regions in the spectra where
all pixel having observed optical depth
larger than 2.5 for \lya transition.
Upward arrows are gaps immediately blueward of quasar
proximity zone, therefore
the length is only a lower limit.
Solid lines with error bars are
average depth lengths with 1-$\sigma$ dispersion at each redshift bin.
Long dark gaps start to appear at $z\sim 5.6$, with the average gap length
increases rapidly at $z>6$, marking the end of reionization.
It is compared with simulation of Paschos \& Norman (2005),
in which dashed and dotted lines are for moderate and high
spectral resolutions.
The simulation has an overlapping redshift at $z \sim 7$.
Adapted from
Fan et al. (2006b).}
\label{gapevo}
\end{figure}

With the emergence of complete Gunn-Peterson troughs at $z>6$, it
becomes increasingly difficult to place stringent limits on the
optical depth and neutral fraction of the IGM.  Songaila \& Cowie
(2002) suggested using the distribution of optically thick, dark gaps
in the spectrum as an alternative statistic.  Fan et al. (2006b)
examined the distribution of dark gaps with $\tau > 2.5$ among their
sample of SDSS quasars at $z>5.7$, and showed a dramatic increase in
the average length of dark gaps at $z>6$ (Figure \ref{gapevo}),
similar to the model prediction of Paschos \& Norman (2005). Gap
statistics provides a powerful new tool to characterize the IGM
ionization at the end of reionization
(Gallerani et al. 2006, Kohler et al. 2006), and can be sensitive to
larger neutral fractions, as they carry higher-order information than
optical depth alone.  Songaila \& Cowie (2002), Pentericci et
al. (2002), Fan et al. (2002, 2006b) also studied using the distribution
function of transmited fluxes (e.g. Rauch et al. 1997) or statistics
of threshold crossing (e.g. Miralda-Escud{\'e} et al. 1996).

\subsection{Estimating the ionization state and the
neutral fraction}

Measurements of Gunn-Peterson optical depth can be used to derive
the ionization state of the IGM, using parameters such
as the mean UV ionizing background, the mean-free-path of UV photons, and
neutral hydrogen fractions.  The evolution of Gunn-Peterson absorption
is usually described in the context of photoionization.  Weinberg et
al. (1997) present the basic formalism to use the IGM optical depth to
measure the cosmic baryon density. McDonald et al. (2000) and McDonald
\& Miralda-Escude (2001) expanded this work to study the evolution of
the ionizing background at $z<5.2$ by comparing the observed
transmitted flux to that of artificial \lya forest spectra created
from cosmological simulations.  Although slightly different in
technical details, a number of works (Cen \& McDonald 2002, Fan et
al. 2002, Lidz et al. 2002, Songaila \& Cowie 2002, Songaila 2004)
followed the same formalism to calculate the evolution of the ionizing
background at $z>5$ and reached consistent results:
from $z\sim 5$ to $z>6$,
there is an order of magnitude decrease in the UV background and the
mean-free-path of UV photons is shown to be $<1$ physical Mpc at $z>6$
(Fan et al. 2006b).  This scale is comparable to the clustering scale
of star-forming galaxies at these redshifts (e.g. Kashikawa et
al. 2005, Hu et al. 2005, Malhotra et al. 2005).  High-redshift star
forming galaxies, which likely provide most of the UV photons for
reionization, are highly biased and clustered at similar physical
scales to the mean free path. The assumption of a uniform UV
background is no longer valid.  There is a clear indication of
a non-uniform UV background at the end of reionization (Fan et
al. 2006b).  However, current observations at $z>6$ are based on a
handful of quasars.  A much larger sample is needed to quantify the
variation of the ionization state.

\begin{figure}
\centerline{\psfig{figure=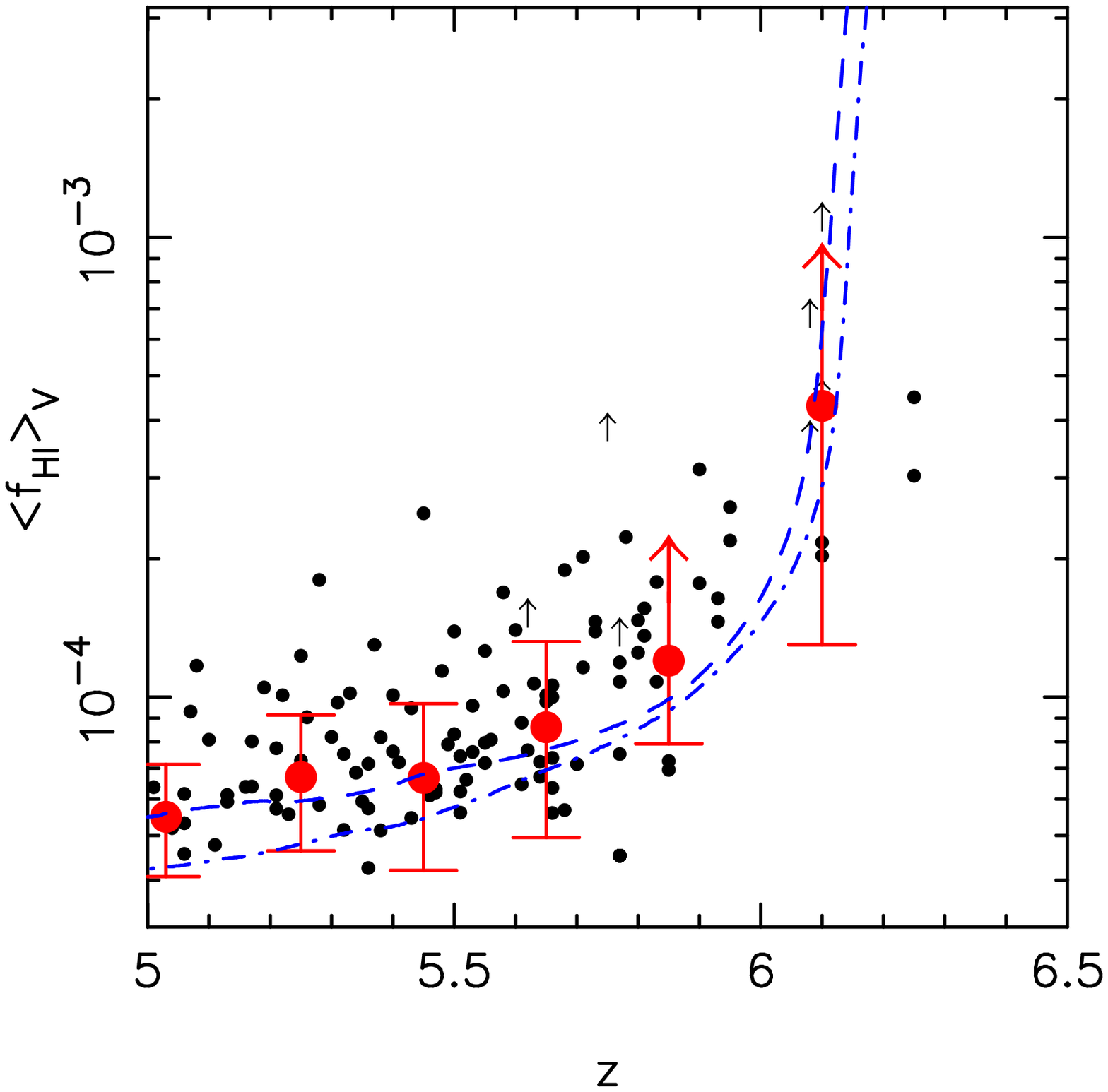,height=30pc}}
\caption{Evolution of volume-averaged neutral hydrogen fraction of the IGM.
The solid points with error bars are measurements based on the
nineteen high-redshift quasars.
The dashed and dashed-dotted-dashed lines is volume-averaged
results from the A4 and A8 simulations of Gnedin (2004).
The neutral fraction inferred from the observations
is comparable to that of the
transition from overlapping stage to post-overlap stage of
reionization in simulations (Gnedin 2000). Adapted from Fan et al. (2006b)}
\label{fHv}
\end{figure}

Using the same data, Fan et al. (2002, 2006b), Lidz et al. (2002), Cen
\& McDonald (2002) find that at $z>6$ the volume-averaged neutral
fraction of the IGM has increased to $> 10^{-3.5}$. The results are
displayed in Figure \ref{fHv}.  It is important to note that this is
strictly a lower limit, due to the large optical depths in the \lya
line.  Further, the presence of transmitting pixels and the finite
length of dark gaps in the quasar spectrum can also be used to place
an independent upper limit on the neutral fraction to be $<10 - 30$\%
(Furlanetto et al. 2004, Fan et al. 2006b). At higher neutral
fraction, the Gunn-Peterson damping wing from the average IGM (see \S
4.4) could wipe out any isolated HII region transmissions.

In summary, the latest work on GP absorption toward the highest
redshift QSOs implies a qualitative change in the nature of \lya
absorption at $z \sim 6$, including: (i) a sharp rise in the power law
index for the evolution of GP optical depth with redshift, (ii) a
large variation of optical depth between different lines of sight, and
(iii) a dramatic increase in the number of dark gaps in the spectra.
The GP results indicate that the IGM is likely between $10^{-3.5}$ and
$10^{-0.5}$ neutral at $z\sim6$.  While saturation in the GP part of
the spectrum remains a challenge, the current results are consistent
with conditions expected at the end of reionization, during the
transition from the percolation, or overlap, stage to the post-overlap
stage of reionization, as suggested by numerical simulations (Paschos
\& Norman 2005; Gnedin 2000, 2004; Ciardi et al. 2001, Razoumov et
al. 2002, Gnedin \& Fan 2006).  Sources at higher redshift, and
estimators sensitive to larger neutral fractions, are needed to probe
deeper into the reionization era.  Wide-field deep near-IR surveys
using dedicated 4-meter class telescopes, such as UKIDSS (Lawrance et
al. 2006) and VISTA (McPherson et al. 2004), will likely discover
quasars at $z\sim 8$ in the next decade (see also \S5.2 for GRB
observations).

\section{Cosmic \strom  spheres and surfaces}

Luminous quasars produce a highly ionized HII region around them even
when the IGM is still mostly neutral otherwise.  The presence of (time
bounded) cosmic \strom spheres (CSS) around the highest redshift SDSS
QSOs has been deduced from the observed difference between the
redshift of the onset of the GP effect and the systemic redshift of
the host galaxy (White et al. 2003; Wyithe \& Loeb 2004a; Wyithe et
al. 2005; Walter et al. 2003).  The physical size of these spheres is
typically $\sim 5$ Mpc at $z>6$, or an order of magnitude larger
volume than the typical spheres expected from clustered galaxy
formation (Furlanetto, Zaldarriaga, \& Hernquist 2004), due to the
extreme luminosity of the QSOs ($\sim 10^{14}$ L$_\odot$).  The size
of the \strom spheres is determined by the UV luminosity of the QSO,
the HI density of the IGM, and the age of the QSO.  Wyithe \& Loeb
(2004a) and Wyithe et al. (2005) use QSO demographics, plus the
distribution of CSS sizes, to infer a mean neutral IGM fraction of
$x_{HI} \ge 0.1$ at $z \sim 6$.

A complementary analysis considers the `cosmic \strom   surface'
around the $z = 6.28$ QSO SDSS J1030+0524.  Using high quality
spectra (see Figure \ref{white1}), Mesinger \& Haiman (2004) show
that the apparent size of the CSS around this QSO is smaller using
the \lya than the \lyb line.  They attribute this difference to
the damping wing of \lya, and from this infer an IGM neutral
fraction $x_{HI}\ge 0.2$.

However, Oh \& Furlanetto (2004) highlight the many significant
uncertainties in both these calculations, including: knowledge of the
systemic redshift of the source, unknown 3D geometry, QSO age
distribution, effect of large scale structure and clustering of
ionizing sources around early luminous quasars, and often noisy
spectra taken in a difficult wavelength range.  In particular,
recombinations in denser regions can be a significant photon sink, and
hence the clumping factor becomes a key parameter. Yu \& Lu (2005)
estimate that these factors could result in up to an order of
magnitude uncertainity in the derived neutral fraction.  Also, there
remains debate as to whether these techniques measure a mass or a
volume averaged neutral fraction.  Fan et al. (2006b) show that over
$5.7 < z < 6.4$, the average size of the quasar \strom spheres
decreases by $\sim 2.5$, consistent with an increase of the neutral
fraction by a factor $\sim 15$ over this redshift range, assuming the
\strom radius $R_s \propto F(\rm HI)^{-1/3}$ and other conditions
being the same.  Measurements of CSS are sensitive to much larger
neutral fraction than using Gunn-Peterson effect, but are also
strongly affected by systematics.  More examples of $z > 6$ QSOs over
a larger range in intrinsic luminosity are required to verify these
calculations.

\section{Other IGM Probes}

In this section, we describe other IGM probes that are either sensitive
to a large neutral fraction, or could be extended to higher redshift in
the near future.

\subsection{Evolution of metal absorption systems}

The early star formation that presumably started cosmic reionization
also enriched the ISM and IGM with heavy elements.  Detailed metal
enrichment models in the early universe are reviewed in Ciardi \&
Ferrara (2005). Songaila (2001, 2005) and Pettini et al. (2003)
found that the total CIV abundance $\Omega(\rm CIV)$ does
not significantly evolve over $1.5 < z < 5.5$.  Schaye et al. (2003)
found a lower bound of [C/H] $\sim -4$ even in the underdense regions
of the IGM at $z<4$.  Madau et al. (2001, see also Stiavelli et
al. 2004) found that by assuming massive stars ionized the Universe
with yield from Population II or III stars, the Universe would have a
mean metallicity of $\sim 10^{-3 \sim -4} Z_{\odot}$ after
reionization, comparable to those found in the high-redshift quasar
absorption lines.  While these measurements do not directly constrain
the reionization history, they suggest that one could detect numerous
metal absorption lines even at $z>6$.

Furlanetto \& Loeb (2003) suggest using observations of low-ionization
absorption lines such as CII or OI to constrain properties of the
stellar population that ionized the Universe at $z>6$, such as star
formation efficiency and escape fraction from supernova winds.  To
directly constrain the reionization history, Oh (2002) suggested using
OI$\lambda$1302 absorption as tracer of the neutral fraction.  OI has
almost identical ionization potentials to H and should be in tight
charge exchange equilibrium with H, while its lower abundance means
that it would not saturate even when the Universe was mostly neutral:
$\tau_{\rm OI}^{\rm eff} = 10^{-6} \left( \frac{\langle Z
\rangle}{10^{-2} Z_{\odot}} \right) \tau_{\rm HI}^{\rm eff}$.  OI (and
SiII$\lambda$ 1260) forests could provide combined constraints on the
reionization and metal enrichment histories. Oh (2002) predicted that
a handful of OI lines could be detected in the Gunn-Peterson trough
redshift regions of known $z>6$ quasars when observed at high
resolution. Becker et al. (2006)
obtained high-resolution,
high S/N spectra of a sample of six quasars at $z>5$ using Keck/HERES
and detected OI system up to $z=6.26$. They do not find a dense OI
forest, consistent with high degree of IGM ionization at $z\sim
6$. However, it is puzzling that the line of sight of SDSS J1148+5251
($z=6.42$), which has the highest ionization fraction at $z>6$
(\S3.3), also has the highest density of OI lines, raising the
possibility that low metalicity, not high ionization, may be the cause
for the lack of OI lines. Detailed modelling of IGM enrichment is
needed to inteprete high-redshift metal line results.

\subsection{GRBs}

Gamma-Ray Bursts (GRBs) are the most powerful explosions in the
Universe, and could be detected at $z>10$. At high-redshift, the time
dilation means that their afterglow will fade away $(1+z)$ times
slower, aiding rapid spectroscopic follow-up observations (e.g. Ciardi
\& Loeb 2000) to probe the IGM evolution.
GRB afterglow has been detected up to $z=6.30$ (SWIFT GRB 050904,
Price et al. 2005, Tagliaferri et al. 2005, Haislip et al. 2005, Kawai et al., 2006).

For a largely neutral IGM, $\tau \sim 10^5$, the damping wing of the
GP trough arising from the large GP optical depth of the neutral
medium will extend into the red side of the \lya emission line
(Miralda-Escud\'e 1998).  For $z\sim 6$, at $\sim 10$\AA\ redward of
Ly$\alpha$ of the host galaxy, the optical depth is of order unity for
a neutral IGM.  However, this GP damping wing test cannot be applied
to luminous quasars, due to the proximity effect from the quasar
itself, as shown by Madau \& Rees (2000) and Cen \& Haiman (2000) and
discussed in \S 4.  Absorption spectra of GRBs, however, are not
affected by the proximity effect and can be used to probe the
existence of damping wings and measure IGM neutral fractions up to
order of unity.  However, strong internal absorption from the neutral
hydrogen of the ISM ($\log$ N(HI) $>$ 21) in the host galaxy appears
to be ubiquitous among GRB afterglow spectra (e.g., Chen et al. 2005).
Such internal absorption, or gas infall in the host galaxy environment
(Barkana \& Loeb 2004), will complicate the interpretation of GRB
observations.  At large distance from line center, the damping wing
from a diffused IGM has a profile $\tau \propto \Delta \nu^{-1}$,
instead of $\tau \propto \Delta \nu^{-2}$ for a discrete absorber.
Totani et al. (2006) fit a \lya absorption profile of GRB 050904
($z=6.30$) with contributions from the internal damped Ly$\alpha$
absorption and the diffuse IGM damping wing simultaneously, and
obtain a conservative limit on the IGM neutral fraction $x_{HI} <
0.6$.

\subsection{Evolution of the IGM thermal state}

\begin{figure}
\centerline{\psfig{figure=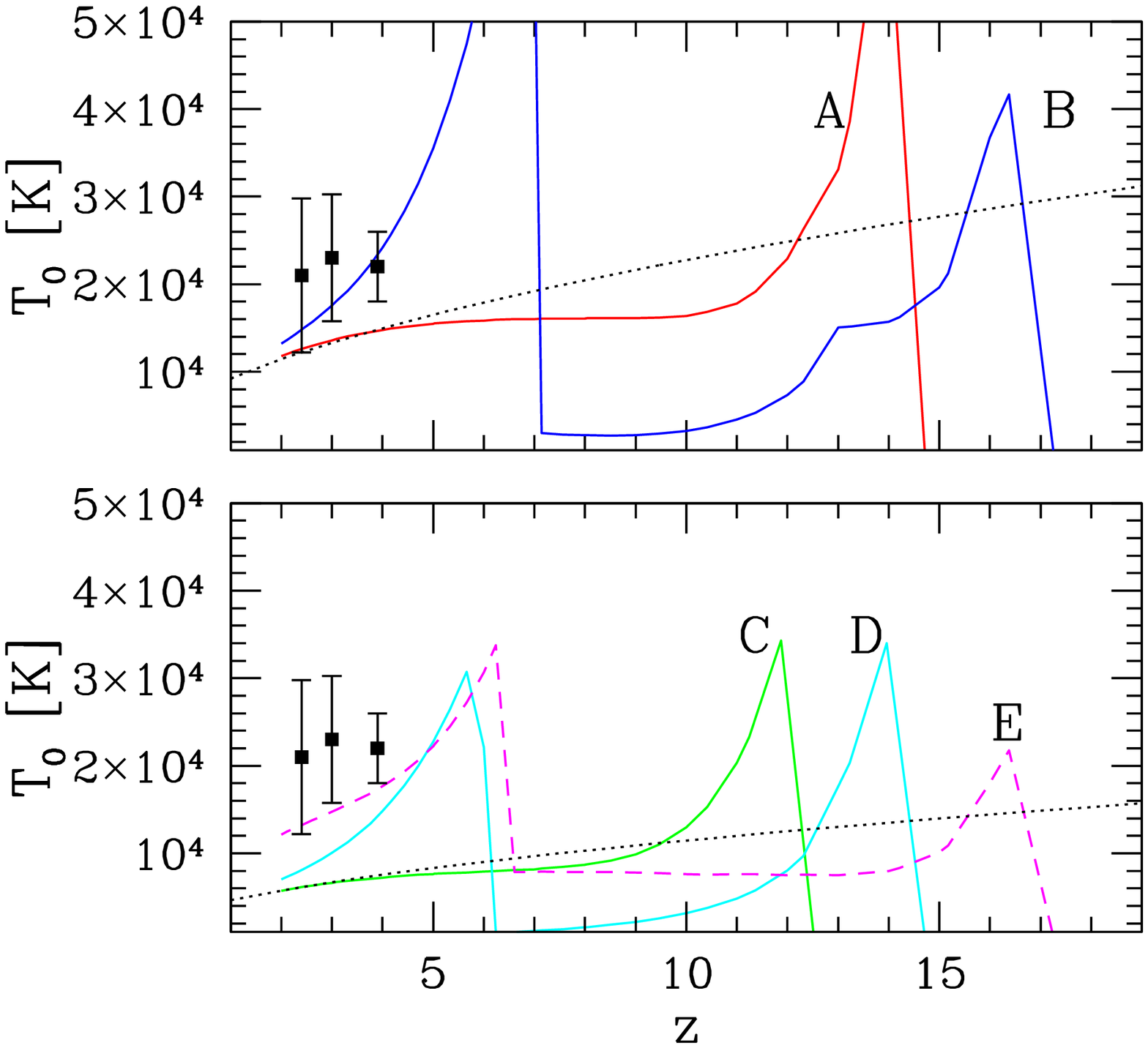,height=30pc}}
\caption{Evolution of IGM temperature for a
quasar-like ionizing spectrum, assuming different
reionization history.
In models where the Universe is reionized early and remained such afterwards,
the IGM temperature is too cold comparing to measurements at $z= 2 - 4$.
Model A has a single-episode reionization at $z=14$ and remains highly
ionized thereafter. Model B also has early reionization, but experienced
a drop in ionizing flux until $z\sim 7$ at which it undergoes a second period
of reionization.
To prevent overcooling, a major heating event that resulted in large
change in the IGM ionization state would have to occur at $z=6 - 10$.
However, different history of HeII reionization history may complicate
this intepretation.
Adapted from Hui \& Haiman (2003).}
\label{thermal}
\end{figure}

Reionization will photo-heat the IGM to several times $10^4$ K.  After
this episode, the IGM will gradually cool mostly due to the Hubble
expansion.  Because of its long cooling time, the IGM will retain some
of its thermal memory of reionization -- earlier reionization leads to
a cooler IGM at lower redshifts.  Theuns et al. (2002a) used absorption
line width measurements to estimate the IGM temperature at $z=2 - 4$,
and found an average temperature of $\sim 2.5 \times 10^{4}$ at $z\sim
3.5$. This temperature constrains reionization to be: $z_{\rm reion} <
9$.  Hui \& Haiman (2003) carried out a similar analysis and
considered different ionizing sources and different ionization
histories.  Figure \ref{thermal} shows their models with different
reionization redshifts and a quasar-like ionizing spectrum.  They
concluded that for all models where the Universe was ionized at
$z>10$, and remained ionized thereafter, the IGM would have reached an
asymptotic temperature too cold compared with observations.  The IGM
thermal history measurement requires the ionized fraction of the IGM
to have of order unity changes at late ($6 < z < 10$) epoch,
regardless of whether there is a very early episode of reionization.
This independent constraint on the reionization history is consistent
with Gunn-Peterson measurements, and it points to a rapid ionization
transition at low redshift.
However, this is a difficult observation, and the interpretation of IGM
temperture evolution may be
further complicated by IGM heating during HeII reonization at lower redshift,
(e.g., Sokasian et al. 2002).

\subsection{Luminosity function and line profiles of Ly$\alpha$ galaxies}

Surveys of galaxies with strong Ly$\alpha$
emission lines  through narrow-band imaging in selected dark windows
of the night sky OH emission forest have proven to be a powerful
technique for discovering the highest redshift galaxies
(e.g., Hu et al. 2002,
2004, Kodaira et al. 2003, Rhoads et al. 2003, Malhotra \& Rhoads
2004, Tran et al. 2004, Kurk et al. 2004, Santos et al. 2004, Martin
\& Sawicki 2004, Taniguchi et al. 2005).  At the time of this writing,
$> 100$ Ly$\alpha$ galaxies candidates have been found at $z\sim 6.5$,
including $>30$ with spectroscopic confirmations, plus the most
distant galaxy with a confirmed spectroscopic redshift, SFJ
J132418.3+271455 at $z=6.589$ (Ajiki et al. 2003).  Currently,
ambitious surveys of Ly$\alpha$ galaxies at even higher redshift,
through windows in the near-IR, are underway (e.g. Barton et al. 2004,
Horton et al. 2004).

Ly$\alpha$ galaxies represent a significant fraction of star forming
galaxies at high redshift (Bouwens et al. 2006).
Properties of Ly$\alpha$ galaxies directly probe the
IGM neutral fraction.  As described above, in a largely neutral IGM,
the Gunn-Peterson damping wing extends to the redside of Ly$\alpha$
emission.  Without a large \strom  sphere, the intrinsic Ly$\alpha$
emission will be considerably attenuated. In the simplest picture, one
predicts: (1) the Ly$\alpha$ galaxy luminosity function will decrease
sharply in an increasingly neutral IGM, even if the total star
formation rate in the Universe remains roughly constant, and (2)
the Ly$\alpha$ profiles will have a stronger red wing
and a smaller average equivalent width before the onset of
reionization.

\begin{figure}
\centerline{\psfig{figure=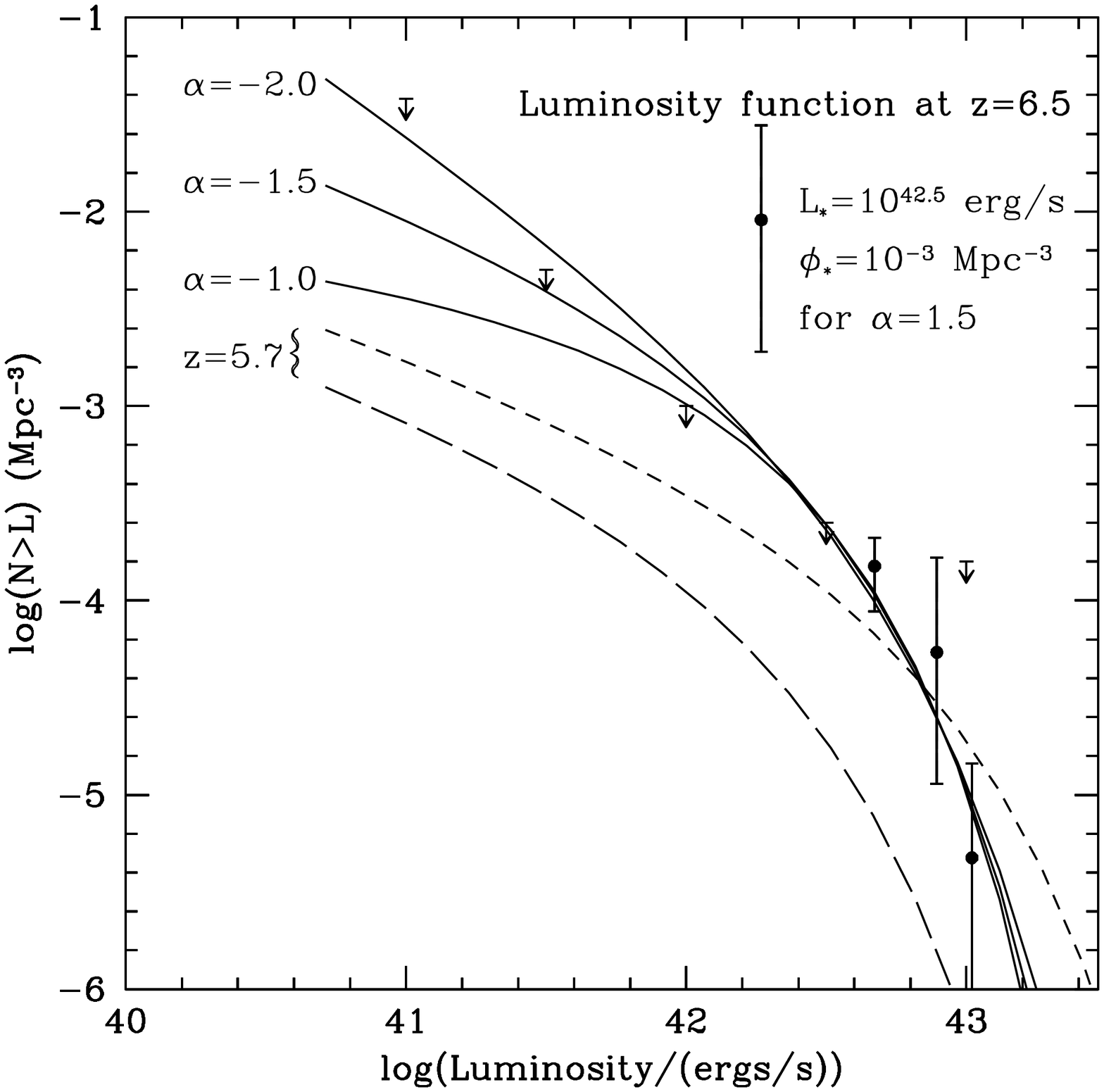,height=30pc}}
\caption{The luminosity function
of Ly$\alpha$ galaxies at $z\sim 6.5$ and $z\sim 5.7$,
from the complications in Malhotra \& Rhoads (2004).
The luminosity function shows little evolution between these
redshifts, consistent with a largely ionized IGM at $z\sim 6.5$.
Adapted from Malhotra \& Rhoads (2004).}
\label{rhoads}
\end{figure}

Malhotra \& Rhoads (2004) and Stern et al. (2004) combined the LALA
survey of Ly$\alpha$ galaxies (Rhoads et al. 2003) with other
Ly$\alpha$ surveys in the literature to determine the luminosity
function of Ly$\alpha$ galaxies at $z=6.5$ and 5.7. They found no
evolution between these two redshift bins,
consistent with the IGM being largely ionized by $z\sim 6.5$ (Figure
\ref{rhoads}).  Hu et al. (2005) constructed Ly$\alpha$ line profiles
at $z=6.5$ and 5.7 from their surveys using Keck and Subaru,
and found a similar lack of evolution.
The interpretations of these results, however, require more detailed
modelling.  Haiman (2002), Santos (2004), Cen et al. (2005) showed
that the local HII regions around Ly$\alpha$ galaxies reduce the
attenuations of Ly$\alpha$ flux.  Gnedin \& Prada (2004), Wyithe \&
Loeb (2005b), Cen (2005) and Furlanetto et al. (2005) further
demonstrated that the clustering of ionizing sources increases the HII
region size and further reduces the attenuation.  Haiman \& Cen (2005)
estimated that the lack of evolution in the Ly$\alpha$ luminosity
function is consistent with the neutral fraction $x_{HI} < 0.25$,
when no clustering is considered
(see also Malhortra \& Rhoads 2005).  Including large scale clustering,
the constraint on the IGM neutral fraction becomes less stringent
(Furlanetto, Zaldarriago \& Hernquist 2005).
Santos (2004) showed that the presence of galactic winds may also play a
crucial role in determining the observed Ly$\alpha$ fluxes.
Haiman \& Cen (2005) suggest studies of
Ly$\alpha$ profiles as a function of luminosity, compared with
low-redshift samples, could provide robust diagnostics to the IGM
ionization state.  Furlanetto et al. (2005) showed that at large
neutral fraction, only galaxies in large ionized bubbles can be
detected in Ly$\alpha$, which have larger bias than individual
galaxies.  With future large area Ly$\alpha$ surveys, it may be
possible to measure the size distribution of HII regions
during reionization in order to constrain reionization history.

\section{Cosmic Microwave Background Probes of Reionization}

The Cosmic Microwave Background (CMB) is a rich ``fossil record''
of the early universe. The CMB not only reveals the universe's
initial conditions, but also its structure and dynamics from the
time of the CMB's origin 400,000 years after the Big Bang to
today. A CMB photon arriving in a terrestrial, balloon-borne, or
space-based CMB observatory is encoded with the conditions of the
universe along its 13.7 billion year journey. In particular, since
the CMB is the oldest observable electromagnetic radiation it acts as a backlight,
illuminating the transition from bulk-neutrality to complete
reionization. Both the temperature anisotropy and polarization of
the CMB help reconstruct the details of the EoR. As we will see,
the CMB's polarization and temperature anisotropy are completely
\emph{complementary} probes; both in their spatial structure, and
in their dependence on the conditions of the EoR which they probe.

\subsection{The Generation of CMB Polarization and Temperature Anisotropy During the EoR}

Thomson scattering produces CMB polarization \emph{only} when
free-electron scatterers are illuminated by an anisotropic photon
distribution \cite{hu1997}. Moreover, the photon anisotropy, when
decomposed into spherical harmonics $Y_{\ell,m}$ in the rest frame of the
electron, must possess a non-zero quadrupole ($\ell=2$) term -- no other 
contributes due to the orthogonality of the spherical harmonics.
Both free-electrons and anisotropically distributed photons were
present during the transition from complete ionization to the EoR.
This epoch produced the primary CMB polarization signal, referred
to as ``E-mode'' or ``gradient mode" owing to its symmetry
properties under parity transformations \cite{zaldarriaga1997a,
kks1997}. The anti-symmetric component of CMB polarization is
referred to as ``B-mode" or ``curl mode" and arises in
inflationary cosmological models which predict a primordial
gravitational wave background. The E-mode polarization anisotropy is produced
by the same perturbations which produce temperature anisotropy and
peaks at $\simeq 10^\prime$ angular scales (or multipoles $\ell
\simeq 1000$) with an amplitude approximately $10\%$ of the CMB
temperature anisotropy at large ($>10\arcdeg$) angular scales.

We begin this section by briefly reviewing how the EoR
produces new CMB polarization and temperature anisotropy. The
physics of CMB polarization and anisotropy has been treated in
numerous sources. We refer the reader to \cite{zaldarriaga1997a,
kks1997, hu1997, dodelson2003} for analytic and theoretical
treatments and \cite{seljak1996} for numerical calculations.
Ultimately, the CMB data, in combination with the 21 cm emission
and Ly$\alpha$ absorbtion by neutral HI, will allow for a detailed
reconstruction of the physics of the EoR.

\subsubsection{Large Angular Scale CMB Polarization and Reionization}

Large-angular scale CMB polarization as a probe of the ionization
history of the universe, has been considered in
e.g. \cite{zaldarriaga1997b, keating1998, kaplinghat2003}. The
importance of large-angular scale CMB polarization is that, unlike
the 21 cm and Ly$\alpha$ HI-absorption measurements discussed
elsewhere in this review, the polarization of the CMB is sensitive
to \emph{ionized} hydrogen (HII) as it is generated by Thomson
scattering.

Reionization produces free-electrons which Thomson-scatter CMB
photons, producing CMB polarization. For angular scales smaller
than the horizon at reionization the scattering damps the CMB
temperature in direction $\hat{\textbf{n}}$ as
$T^\prime(\hat{\textbf{n}})=e^{-\tau}T(\hat{\textbf{n}}),$ where
$\tau$ is the Thomson optical depth. Reionization, therefore,
damps the temperature anisotropy power angular spectrum as
$C_\ell^{T^\prime}=e^{-2\tau}C_\ell^T$. The damping of the
temperature power spectrum is degenerate with the primordial power
spectrum's amplitude, $A$, for scales smaller than the horizon at
last scattering.

\begin{figure}[t]
\centerline{\psfig{figure=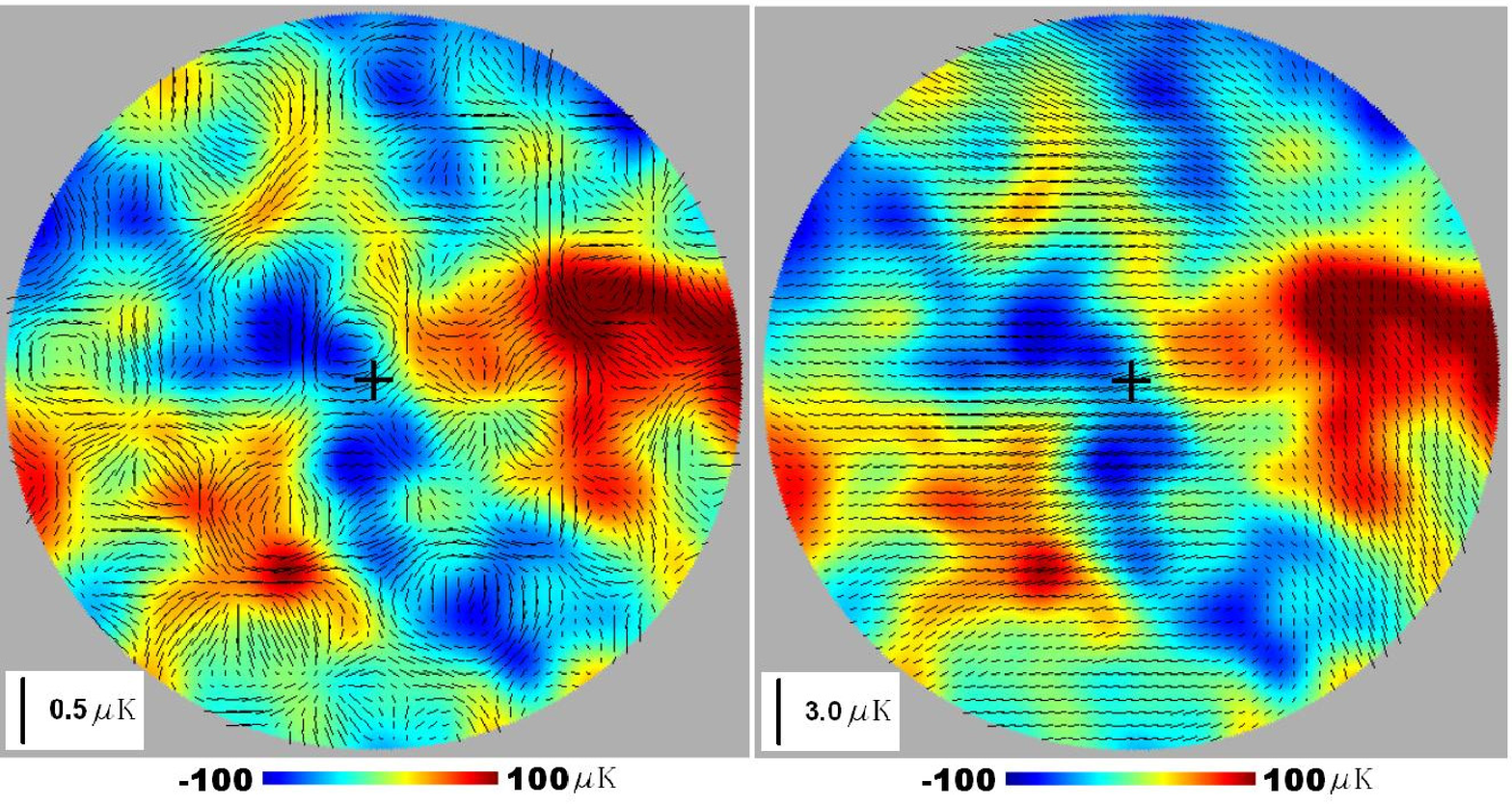,angle=90,height=8cm}}
\caption{\label{fig:polmaps} Large angular scale temperature
(color scale) and polarization (lines) simulations in real-space.
The left panel shows a simulation with no reionization and the
right panel shows the effects of instantaneous reionization with
$\tau=0.17$. Both maps and CMB observables are smoothed with a
$4\arcdeg$ beam. Reionization does not affect the large-scale
temperature pattern but it produces roughly ten-times larger
polarization at large scales -- note the change in scale for
polarization between the two panels. The maximum polarization in
the right panel is $\simeq 800$ nK. Figures courtesy of Eric
Hivon.}
\end{figure}

Fortunately, a new feature, e.g. \cite{zaldarriaga1997b}, in the
E-mode power spectrum develops at large angular scales which
breaks the degeneracy between $A$ and $\tau$. The degeneracy (for
the CMB temperature anisotropy) and its breaking (using CMB
polarization) is illustrated in (real-space) in figure
\ref{fig:polmaps}. Figure \ref{fig:emode_limits} shows the
polarization angular power spectrum associated with these
simulations, along with current observational results (discussed
further in \ref{subsec:results}). Note in \ref{fig:emode_limits}
the new large-angular scale (low-$\ell$) peak which discriminates
between models with and without reionization.

The polarization anisotropy of the CMB reveals the ionization
condition of the universe in ways which the temperature anisotropy
cannot. Following recombination at $z = 1089 \pm 1$
\cite{spergel2003} primordial hydrogen remained neutral until the
epoch of reionization. The term ``last-scattering" is only
approximately true since we know that the universe is nearly 100\%
ionized at present ($z=0$). Therefore some fraction (few to $\sim
10\%$) of the photons scattered \emph{again} following decoupling.
If reionization was instantaneous, only the primordial temperature
quadrupole at the ``last-scattering surface" ($z\simeq 1100$)
projected to the redshift of reionization contributes to the CMB
polarization at the $< 1\muK$ level (at $>10\arcdeg$ scales).
This reionization feature has been
detected by WMAP and is interpreted as a detection of $\tau = 0.10
\pm 0.03$, or $\tau = 0.09 \pm 0.03$ when all six cosmological
parameters are fit to all of the WMAP data
(TT,TE,EE)\cite{page2006} .

Whereas CMB temperature anisotropy is generated by processes
occurring after reionization (e.g., gravitational redshifts due to
matter inhomogeneity along the line of sight -- the Integrated
Sachs-Wolfe (ISW) effect), the large-scale polarization of CMB is
not similarly affected. The large-scale CMB polarization is a
cleaner and more precise probe of the onset of the reionization
epoch than the anisotropy for two reasons. The first is that the
ratio of the polarized intensity to the total intensity is
unchanged when the CMB propagates through an absorbing medium
(such as the post-reionization intergalactic plasma) -- both
polarization components are attenuated equally. The second is that
both polarization components incur equal gravitational redshifts
(blueshifts) when climbing out of (falling into) time-varying
gravitational potential wells from last-scattering to today since
gravity cannot distinguish between the two linear polarization
states. This is equivalent to asserting that CMB polarization does
not experience the ISW effect, and so the low-multipole E-mode
spectrum is not contaminated by the ISW effect.

High precision observations of the large-scale polarization will
reveal more of the EoR's details. The polarization power spectra are
sensitive to the duration of reionization.  With upcoming
observations the dynamics of the EoR can be constrained,
complementing 21 cm observations (\S 8),
by providing (modest) redshift information; most notably regarding
the transition from partial to total reionization by constraining
$x_{HI}(z)$ -- difficult using 21 cm data alone. This would be a
powerful probe of ``double reionization" models\cite{cen2003}. A
cosmic variance limited experiment, with sufficient control over
instrumental systematics and foregrounds, can distinguish the
transition from partial to total reionization using details of the
polarization power spectra \cite{kaplinghat2003,holder2003}.
Kaplinghat \textit{et al.} show that large-scale CMB polarization can
constrain the redshift of the transition from partial to total
reionization with precision
$\sigma_{z_{ri}}=333\sigma_\tau/\sqrt{z_{ri}}$, where
$\sigma_\tau$ is the precision with which the experiment measures
the optical depth. Upcoming measurements, such as Planck, will
produce percent-level constraints on partial reionization or
double reionization scenarios and discriminate between models with
identical $\tau$, but very different ionization
histories\cite{kaplinghat2003}.

Reionization also produces a new low-$\ell$ peak in the B-mode
polarization pattern, which is absent in models without
reionization. Even a modest ionization fraction increases the
observability of the B-mode (curl-mode) polarization. This is
significant because gravitational lensing of large scale structure
converts E-mode power into B-mode power \cite{zaldarriaga1998,
huokamoto2002}. This contaminant increases with increasing
multipoles up to $\ell \sim 1000$. With the new, low-$\ell$
structure, a more stringent limit on the inflationary-generated
gravitational wave background can be obtained than would be
possible without reionization\cite{kaplinghatknoxsong2003}.

\subsubsection{Small Angular Scale CMB Polarization and Reionization}
Because even the primary CMB polarization is so small, observing
effects to second order in $\tau$ or distinguishing the details of
the Str\"{o}mgrensphere percolation process is extremely
challenging using CMB polarization.
Fortunately, the small angular scale CMB \emph{anisotropy} will provide a
rich data set with vital information on the fine-details of the
reionization epoch.

\subsubsection{The effects of reionization on CMB temperature anisotropy}

CMB polarization probes the onset of reionization and
distinguishes partial reionization from complete, but it cannot
probe the details of the ionization percolation process as it is sensitive to the integrated Thomson optical depth to the last scattering surface. In fact, the reionization details which \emph{can} be provided by large
scale polarization is known to be limited by cosmic variance
\cite{HuHolder03} and the smoothing inherent in large scale
polarization measurements. Fortunately, the CMB temperature
anisotropy at small angular scales may reveal many of these
interesting features \cite{Aghanim1996,gruzinovhu1998,knox1998,
haiman1999, BL01, santos2003, zahn2005}. Thus, both the bulk and
fine details of reionization will be probed by the complementary
measurements of large-scale CMB polarization anisotropy and
small-scale temperature anisotropy observations.

In non-instantaneous models of reionization (due to discrete
ionizing sources), reionization proceeds in a ``patchy" manner
where the Str\"{o}mgrenspheres of ionized hydrogen surrounding the
sources eventually coalesce to produce the fully-ionized universe
observed out to $z \sim 6$. This patchiness induces secondary CMB
temperature anisotropy
\cite{Aghanim1996,gruzinovhu1998,knox1998,mcquinn2005}. In
addition to the damping of the CMB temperature anisotropy power
spectrum, there are two interesting \emph{secondary} temperature
anisotropy effects: 1) the kinetic Sunyaev-Zel'dovich effect (kSZ)
and 2) the Ostriker-Vishniac effect (OV).

The kSZ effect is due to the motion of regions of reionized
electrons along the line of sight. It is similar to the kinematic
SZ effect produced by inverse Compton scattering of the CMB by
electrons in moving galaxy clusters \cite{sunyaev1980}, but here
it refers to the motions of the ionized electrons in the
pre-galactic medium (PGM) during the EoR. At very small angular
scales ($<4^\prime$) the primary CMB anisotropy signal is damped
due to photon diffusion out of overdense regions (Silk damping)
prior to decoupling. Here the thermal SZ effect dominates over the
primary CMB anisotropy, but at the thermal SZ null (218 GHz) the
EoR kSZ effect may be the dominant source of CMB temperature
anisotropy.

During reionization, electron spatial inhomogeneity can arise in two
ways: either by a variable ionization fraction, $x_i$, of by a
\emph{constant} $x_i$ and a \emph{variable} baryon density. This
effect is the OV effect which is primarily generated by structures in
the non-linear regime, \emph{i.e.}, at low-$z$. The OV effect is the
result of a ``patchy" reionzed medium. Much progress has been made due
to the development of fast structure formation codes (e.g., smooth
particle hydrodynamics -- SPH) which trace the reionization mechanism
(since they trace the formation of baryonic structures).  The
reionization efficiency of the baryonic structures is the only free
parameter, and many of these approaches, e.g., Zahn et al.  (2005)
incorporate semi-analytic reionization models into their
simulations. Using an extended Press-Schechter formalism to model the
reionization process Zahn et al. (2005) find that patchy reionization
makes a large contribution to the CMB anisotropy at scales of
$6^\prime$ ($\ell \sim 2000$) which exceeds the kSZ effect. In fact,
ignoring these secondary, EoR effects will lead to significant biases
in the estimation of cosmological parameters derived from the primary
CMB anisotropy, which further underscores vital importance of their
measurement \cite{santos2003, zahn2005}.

Both the kSZ effect and the OV effects occur at small scales
($<0.1\arcdeg, \ell >2000$) and probe both the
homogeneity and efficiency of the reionization process. The shapes
of the predicted temperature anisotropy power spectra for these
secondary effects appear to be robust to changes in the
reionization model \cite{knox2003,santos2003,zahn2005}, though the
peak-position of the secondary power spectra occurs when
$x_{HI} \sim 0.5$, which \emph{is} model-dependent
\cite{FMH05}. In contrast, the overall amplitude of the
secondary spectra appears to be significantly model-dependent.
However, the amplitude depends only weakly on the epoch of
reionization and more strongly on the duration of the patchy
phase, making it complementary to the CMB polarization
measurements of the optical depth\cite{mcquinn2005}. For
completeness we close this subsection on secondary effects by
mentioning that both the \emph{polarized} kSZ and OV effects are
expected to be nearly negligible and practically impossible to
measure\cite{seshadri1998,Hu2000}

\subsection{Current CMB Results and Future Prospects}
\label{subsec:results}

The first detection of reionization phenomena using CMB
observations came with NASA's Wilkinson Microwave Anisotropy Probe
(WMAP) which reported a nearly $5\sigma$ detection of a
reionization feature in the large angular scale ($>10\arcdeg$)
temperature-polarization cross-correlation function ($\langle TE
\rangle$) with only a single year of data\cite{kogut2003}. This
detection was the first use CMB polarization to measure the
physics of the EoR.  It also was the first \emph{detection} of the
Thomson optical depth, $\tau$, rather than the lower limits
(provided by Ly$\alpha$ measurements) or upper limits (provided by
CMB temperature anisotropy measurements). WMAP's initial
detection of $\tau$ in 2003 using the TE-data and subsequent detection
using the E-mode polarization only \cite{page2006}
also  determines the primordial power spectrum amplitude $A$, $\sigma_{A} \simeq
2\sigma_\tau$ \cite{spergel2003}. WMAP's detection illustrates CMB
polarization's power to probe vitally important cosmological
parameters that are essentially unobservable using temperature
anisotropy alone.

WMAP's three-year data set \cite{page2006} reports several
detections of the E-mode and temperature-polarization
cross-correlation power spectrum, especially at low-$\ell$ (due to
WMAP's ability to map most of the sky). CBI, DASI, and
BOOMERANG\cite{piacentini2005} have also detected the $\langle
TE\rangle$ cross-correlation spectrum, at smaller angular scales
than WMAP. WMAP's first-year results are relatively robust to the
choice of additional data sets and priors used to constrain
cosmological parameters (including $\tau$), such as SDSS
(\cite{tegmark2004}) and Ly$\alpha$ forest measurements
\cite{seljak2005}.

\begin{figure}[htb]
\centerline{\psfig{figure=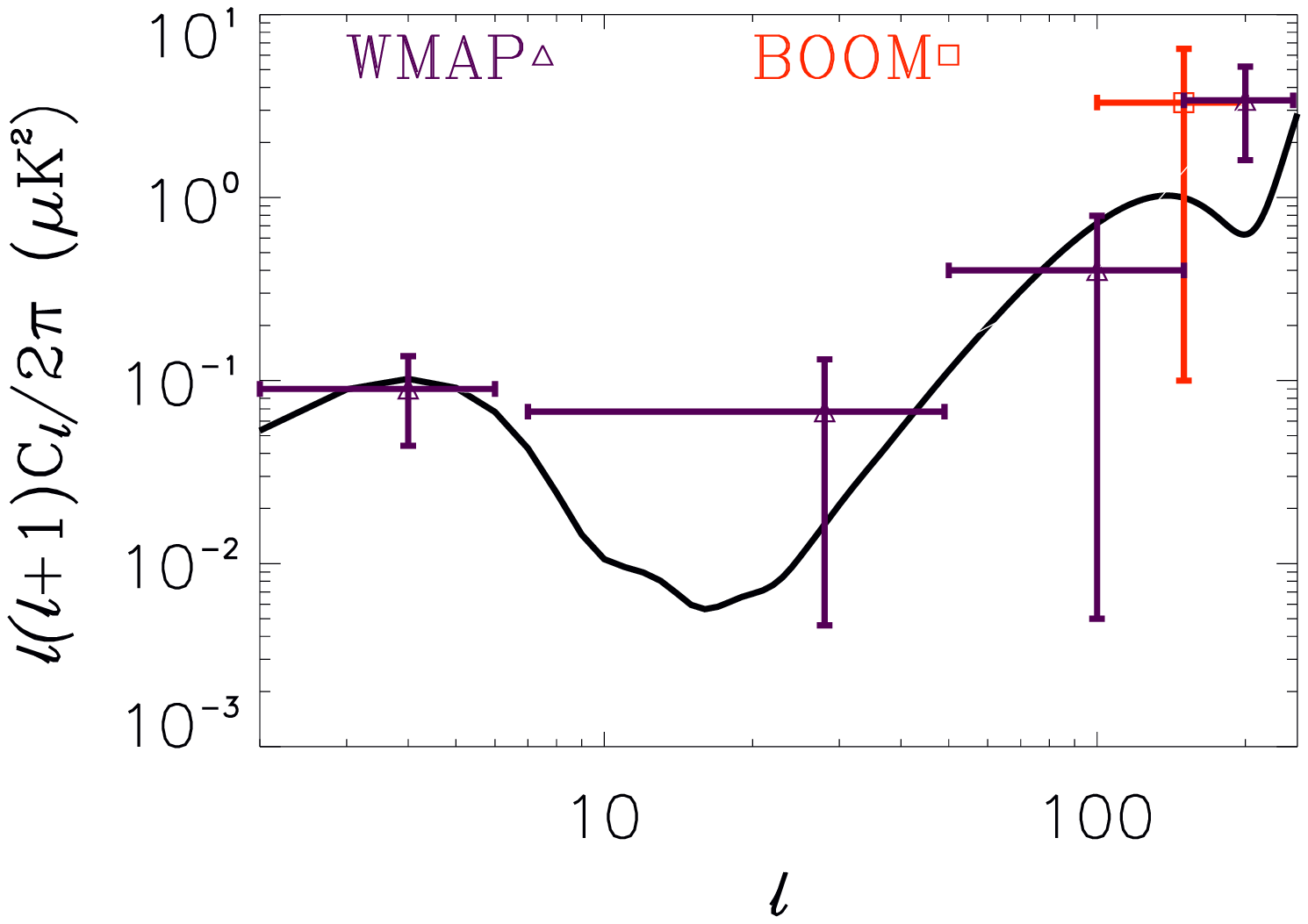,height=30pc}}
\caption{Measurements of the large-angular scale gradient-mode
polarization power spectrum $C_{\ell}^{E}$ for angular scales greater than $1\arcdeg
\Leftrightarrow \ell \lesssim 200$. The solid line is the polarization
power spectrum for the WMAP best fit cosmological model, with
$\tau=0.09$ \cite{spergel2006,page2006}
The dashed line displays the polarization power spectrum for a
model generated using the WMAP cosmological parameters but with
$\tau = 0$, which is consistent with the E-mode detections by
BOOMERANG, CBI, DASI, and CAPMAP and the upper limit of POLAR.}
\label{fig:emode_limits}
\end{figure}

\begin{figure}[htb]
\centerline{\psfig{figure=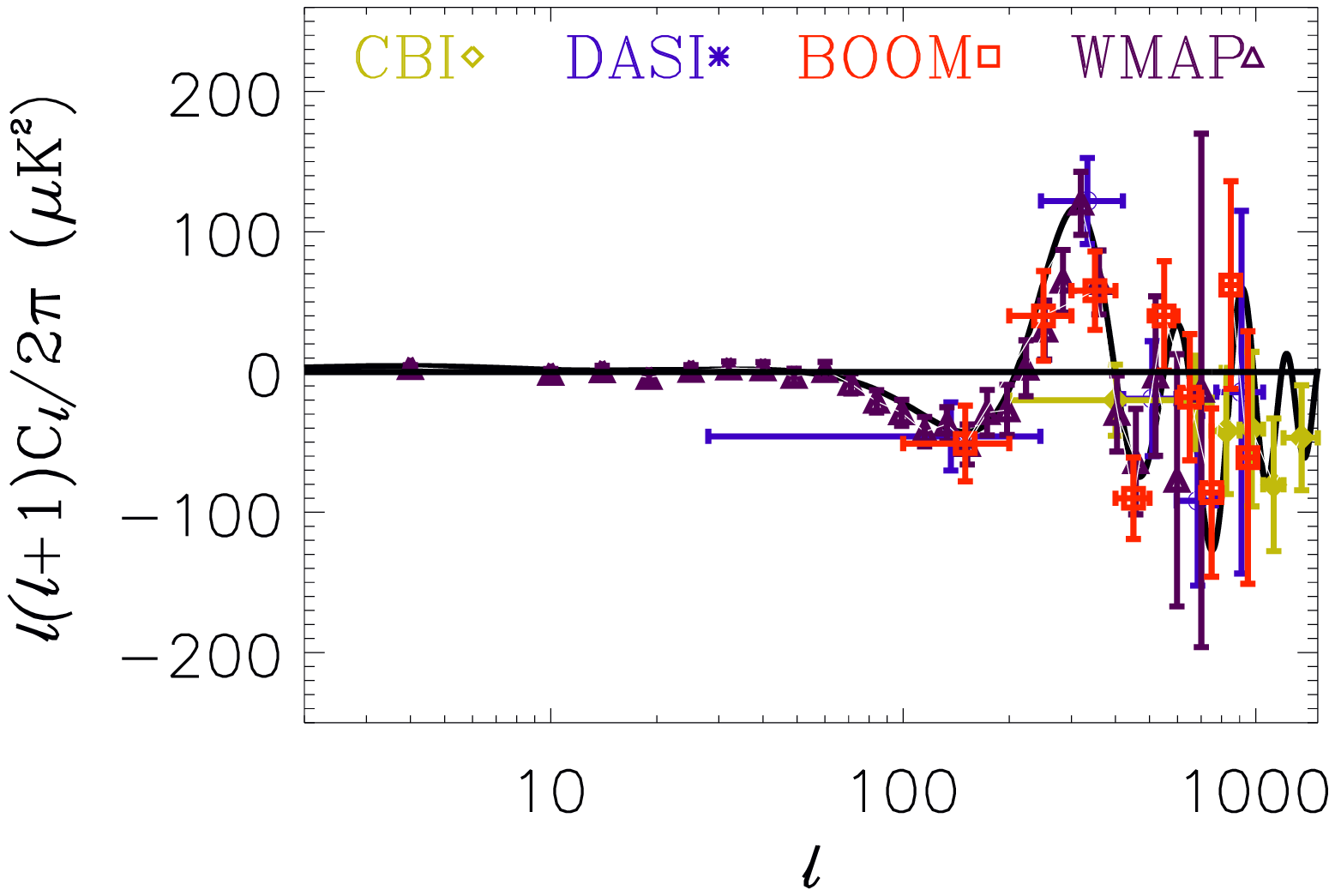,height=30pc}}
\caption{Measurements of the temperature-polarization
cross-correlation power spectrum $C_{\ell}^{TE}$. The solid line
is the power spectrum for the WMAP best fit cosmological model,
with $\tau=0.09$  \cite{spergel2006,page2006}.}
\label{fig:temode_limits}
\end{figure}

In figures \ref{fig:emode_limits} and \ref{fig:temode_limits}
observational CMB polarization and temperature-polarization
cross-correlation power spectra are shown for the angular scales
relevant for probing the EoR. The primary effects of reionization
are encoded in the E-mode autocorrelation and temperature
polarization cross-correlation power spectra at $\ell \lesssim
50$. As demonstrated in, e.g., \cite{zaldarriaga1997b,HuHolder03,
keating2005} essentially all the information on $\tau$ comes from
$\ell<50$. While most of the $\tau$-constraint comes from
$\ell<10$, a non-negligible amount comes from $10<\ell<40$.
Currently, only WMAP\cite{page2006} and
BOOMERANG\cite{montroy2005} have detected E-mode polarization at
angular scales greater than $1\arcdeg$ ($\ell < 200$), and only
WMAP has detected the E-mode polarization for $\ell < 100$.
CBI\cite{readhead2004} and CAPMAP\cite{barkats2005} have detected
the E-mode polarization, though at smaller angular scales (higher
$\ell$) than are relevant for probing the EoR with CMB
polarization alone.

While no detections of the kSZ or OV effects have been reported,
several groups have published detections of CMB anisotropy at
scales which are relevant to both secondary temperature effects.
For $3000 < \ell < 10000$, the kSZ temperature anisotropy power
spectrum scales roughly as $C_l ~ (\Omega_b h)^2 \sigma_8^{p}$
where $3<p<7$ (see \emph{e.g.} \cite{zhangpentrac04}).
\cite{dawson2002} report a detection of CMB anisotropy at
sub-arcminute ($\ell \simeq 7000$) scales using the BIMA array and
\cite{readhead2004b} report a detection at larger (few-arcminute)
scales using CBI. The results of the two groups
are consistent, and seem to indicate significant power in excess of the
CMB primary anisotropy at those scales. Both groups speculate on
the likelihood that this excess structure is due to the thermal SZ
effect. While neither group has detected the kSZ or OV effects,
they are valuable technological and methodological precursors for
future CMB anisotropy measurements of the EoR.

\subsection{Observational Challenges}

A hallmark of CMB experiments since the detection of the CMB has
been the mitigation and excision of systematic effects. CMB
observations have gone to great lengths to combat systematic
effects, including building duplicate telescopes (e.g. the COBE
and WMAP satellites). Another method of improving experimental
fidelity is to conduct extremely wide-band observations for
foreground monitoring and removal. For example, WMAP has
radiometers covering more than two octaves in
frequency\cite{jarosik2003}, and Planck's radiometers will cover
nearly five octaves\cite{lamarre2003,mennella2004}.

Particularly for the large scale E-mode polarization, measuring
the CMB with multiple frequency channels alone is not sufficient.
To overcome statistical (cosmic) variance on the low-$\ell$
$C^E_\ell$ peak, an experiment must observe a large survey region
-- at least $\sim 30\arcdeg \times 30\arcdeg$. Such a large area
is likely to be contaminated by galactic emission, which can only
be subtracted to finite precision. The most conservative approach
to deal with galactic foregrounds is to exclude regions of strong
foreground emission (such as the galactic plane) from the survey.
For example, the WMAP team used multiple frequency channels in
combination with sky-cuts that removing $15\%$ of the sky
surrounding the galactic plane \cite{kogut2003} and introduced
correlations between multipoles at the few percent-level. Even
with sophisticated foreground modelling and multiple frequency
coverage the challenges of achieving polarization fidelity at the
100 nK level over $>10\arcdeg$ angular scales -- required, for
example, to detect the signatures of non-instantaneous
reionization scenarios (e.g, \cite{kaplinghat2003,holder2003}) --
are daunting. The CMB community is well aware of the challenge;
see for example, the CMB Task Force Committee
report \cite{weiss2005}.

Observational challenges for secondary temperature anisotropy (kSZ
and OV) measurements include point sources, primary CMB confusion,
secondary CMB confusion due to gravitational lensing and the
thermal SZ effect. To detect the secondary effects, the primary
CMB anisotropy must be overcome by measuring at scales smaller
than the Silk damping scale, requiring large ($>5$ m class)
telescopes. Of the foregrounds for the kSZ and OV effects, the
thermal SZ effect is perhaps the most straightforward to mitigate
-- by measuring at the thermal SZ-null near 218 GHz. At small
scales, point sources need to be excised based on ancillary
measurements of their spectra and position. There are also
significant hurdles to overcome in the theoretical modelling of
the secondary temperature effects. In particular, separating the
low-$z$ OV effect from high-$z$ patchy reionization effects will
be challenging, though hopefully amenable to theoretical modelling and simulation.

Since galactic foregrounds and point sources contribute to
spurious polarization and anisotropy with different systematics,
combining the large angular CMB polarization observations with the
small scale temperature anisotropy measurements represents the
most promising avenue towards a faithful reconstruction of the EoR
using the CMB.

\subsection{Discussion and Future Prospects}

WMAP's on-going measurements of CMB polarization are expected to
continue until at least 2007. Planck is expected to be launched in
2008 and will start making full-sky observations of the
polarization and primary temperature anisotropy of the CMB. Planck
is expected to achieve a nearly cosmic variance limited
measurement of $\tau$. A balloon-borne CMB polarimeter called EBEX
\cite{oxley2005} will cover a large region of the sky in an
attempt to measure the signature of gravitational waves imprinted
on the CMB.

More reionization information will emerge from CMB B-mode
polarization observations, which search for the imprint of
primordial gravitational radiation on the CMB. Like the E-mode
polarization reionization signature, the B-mode polarization peaks
at super-degree scales so the upcoming experiments optimized to
search for the B-mode signal will have appreciable sensitivity to
reionization features in the E-mode spectrum \cite{keating2005}.

An ancillary benefit of reionization is that it boosts the primary
(inflation generated) B-mode power spectrum significantly near
$\ell = 10$. Due to reionization, a tighter limit on the
tensor-to-scalar ratio, $r$, in the presence of gravitational
lensing (at redshifts $z < 10$) can be obtained than that
calculated in \cite{knox2002,kesden2002}. Kaplinghat, Knox \& Song
\cite{kaplinghatknoxsong2003} demonstrate that the minimum
detectable inflationary energy scale behaves as $1/\tau^4$ due to
the low-$\ell$ B-mode power spectrum peak produced by
reionization. This motivates very wide-field, or full-sky,
observations of CMB polarization. NASA's Beyond Einstein
initiative features a CMB polarimeter called ``CMBPol"\footnote{
http://universe.nasa.gov/program/inflation.html}. This instrument
is designed to detect the signature of inflationary gravitational
waves over a wide range of inflationary energy scales. This
experiment will achieve cosmic variance limited precision on
$\tau$, which is essentially the same as Planck's sensitivity
\cite{kaplinghat2003}, but in contrast to
Planck, CMBPol will be optimized to detect CMB polarization.

Three large-format bolometric array telescopes are currently being
developed to probe CMB temperature anisotropy and the kSZ and OV
effects. The Atacama Cosmology Telescope (ACT,
\cite{kosowsky2003}) is a 6 m telescope operating in three
frequency bands (145, 225, and 265 GHz) with $\simeq 1^\prime$
angular resolution.  With concordance cosmological
parameters, ACT should be able to constrain the redshift of
reionization to percent-level accuracy \cite{zhangpentrac04}. The
South Pole Telescope (SPT, Ruhl et al. 2004) is a 10 m telescope
operating in 5 bands from 100 to 345 GHz with angular resolution
ranging from $1.5^\prime$ to $\simeq 0.5^{\prime}$. Finally, the
largest CMB anisotropy telescope which will be sensitive to
secondary CMB temperature anisotropy is the 12 m Atacama
Pathfinder EXperiment (APEX) which will use a 320 pixel array of
transition edge sensor (TES) bolometers, also from a high altitude
(~5000 m) observatory in the Chilean Atacama desert. All of these
instruments, with the exception of CMBPol, are scheduled for
``first-light" before 2010.

This section has emphasized the role of the CMB in illuminating
the physics of the EoR. Both the CMB's polarization and its
temperature anisotropy provide a wealth of reionization data in a
fashion that is complementary to the 21 cm and Ly$\alpha$
observations discussed elsewhere in this review. While the
importance of large-scale CMB polarization to reionization was
recognized early-on, its ability to reveal more than the bulk
Thomson optical depth was not initially appreciated. Now we
understand that the polarization and the temperature anisotropy
provide a detailed view of the EoR. In fact, they do so in
completely complementary ways: the polarization reveals the epoch
of reionization in a way the temperature anisotropy cannot.
Likewise, the CMB's temperature anisotropy probes secondary
effects to which the polarization is blind. When the new CMB
polarization and temperature anisotropy observations are combined
a high-fidelity image of the EoR will emerge.

\section{Sources of Reionization}

Regardless of the detailed reionization history,
the IGM has been almost fully ionized since at least $z\sim 6$.
This places a minimum requirement on the emissivity of
UV ionizing photons per unit comoving volume required to keep
up with recombination and maintain reionization
(Miralda-Escud{\'e} et al. 2000):
\begin{equation}
\dot{\cal N}_{ion}(z) = 10^{51.2}\,{\rm s^{-1}\, Mpc^{-3}}\left({C\over 30}\right)\times
\left(\frac{1+z}{6}\right)^3 \left({\Omega_b h^2\over 0.02}\right)^2,
\end{equation}
where $C \equiv\frac{\langle n_H^2 \rangle}{\langle n_H
\rangle^2}$ is the clumping factor of the IGM.
It is difficult to determine $C$ from observations,
and has large uncertainty when estimated from simulations
($C = 10 - 100$, Gnedin \& Osteriker 1997).
At $z<2.5$, the UV ionizing background is dominated by quasars and AGN
(e.g. Haardt \& Madau 1996). At $z>3$, the density of luminous quasars
decreases faster than that of star-forming galaxies, and the ionizing
background has an increased contribution from stars (e.g. Haehelt et al. 2001).
In this section we summarize what is known about the ionizing background
contribution from quasars and AGN, young stars, or other sources of
high energy photons (e.g. particle decay).

\subsection{Quasars and AGN}

Quasars and AGN are effective emitters of UV photons. The UV photon
escape fraction is generally assumed to be of order unity.  The UV
photon emissivity from quasars and AGN can be directly measured from
integrating quasar luminosity functions at high-redshift.  The density
evolution of luminous quasars ($M_{1450} < -27$) has been
well-determined from surveys such as SDSS (e.g. Warren et al. 1994,
Kennefick et al. 1995, Schmidt et al. 1995, Fan et al. 2001a, b, 2004,
Richards et al. 2005) up to $z\sim 6$.  Luminous quasar density
declines exponentially towards high-redshift: it is $\sim 40$ times
lower at $z \sim 6$ than at its peak at $z\sim 2.5$.  However, quasars
have a steep luminosity function at the bright end -- most of the UV
photons come from the faint quasars that are currently below the
detection limit at high-redshift.

Fan et al. (2001b) showed that quasars could not have maintained IGM
ionization at $z\sim 6$, as far as the shape of the luminosity
function at $z\sim 6$ is not much steeper than that at $z<3$.  In
fact, Fan et al. (2001a) and Richards et al. (2005) used the SDSS
sample to find that at least the bright end slope of quasars at $z>4$
are considerably flatter than that at low-redshift, consistent with
the findings of the COMBO-17 survey (Wolf et al. 2003).
Miralda-Escud\'{e} (2003), Yan \& Windhorst (2004a) and Meiksin (2005)
used different parameterizations of quasar luminosity function
evolution and came to the same conclusion.  Willott et al. (2005) and
Mahabal et al. (2005) present surveys of a few deg$^2$ for faint
$z\sim 6$ quasars. Their detection of only one $z \sim 6$ faint QSO
implies that the quasar population contribution to the ionizing
background is $<30$\% that of star-forming galaxies.

\subsection{Lyman Break galaxies}

Due to the rapid decline in the AGN populations at very high $z$,
most theoretical models assume stellar sources reionized the universe.
However, despite rapid progress, there is still considerable
uncertainty in estimating the total UV photon emissivity of star-forming
galaxies at high-redshift.
Modifying the constraint from Equation 4,
Bunker et al. (2004) and Bouwens et al. (2005) showed:
\begin{equation}
\dot{\rho_{*}} \approx (0.026 \,\sfrd)\,\left(\frac{1}{f_{\rm
esc,rel}}\right)\,\frac{C}{30} \left(\frac{1+z}{7}\right)^{3}.
\end{equation}
where $\dot{\rho_{*}}$ is the UV star formation rate rate, $C_{30}$ is
the IGM clumping factor,
$f_{\rm esc,rel}$ is the fraction of ionizing radiation
escaping into the intergalactic medium to that escaping in the
$UV$-continuum ($\sim1500\AA$).
The uncertainties come from three factors:
the star formation rate, clumping factor, and UV escape factor.
Current estimates of the star formation rate at $z\sim 6$
all come from photometrically-selected $i$-dropout objects
selected in a small number of deep fields observed
by the HST: the HUDF (Beckwith et al. 2005 in preparation), GOODS fields
(Giavalisco et al. 2004), and UDF ACS Parallels
(Thompson et al. 2005). These estimates are affected strongly by sample
(or cosmic) variance at $z\sim 6$ due to the small volume of deep surveys,
as well as estimates or extrapolations towards faint luminosities.
The clumping factor is typically taken from cosmological simulations
(e.g.  $C\sim 30$, Gnedin \& Osteriker 1997).
The UV escape fraction remains problematic.
Direct measurements of the escape fraction range from
upper limits, $<0.1$ to $<0.4$ (e.g.
Giallongo et al. 2002, Fern{\'a}ndez-Soto et al. 2003,
Inoue et al. 2005) to $>0.5$ (Steidel et al. 2001).
Therefore, instead of measuring the accurate UV photon emissivity,
most efforts have been trying to determine whether star-forming
galaxies could provide sufficient photons to at least meet the
requirement in Equation 5 at $z\sim 6$.

Using HUDF data, Yan \& Windhorst (2004b) found 108 plausible
candidates at $5.5 < z < 6.5$ down to $m_{\rm AB}(z) = 30.0$ mag.
They estimated a steep faint-end luminosity function with a
power-law index of $-1.8$ to $-1.9$, and concluded that this steep
slope is sufficient to satisfy the reionization requirement, and that
most of the photons that ionized the universe come from dwarf galaxies.
Similarly, Stiavelli et al. (2004) compared GOODS and HUDF
observations and found that the observed mean surface brightness of
galaxies at $z\sim 6$ is sufficient for reionization when the young
stellar population has a top heavy IMF.
Bunker et al. (2004), however, found a much lower star formation density
at $z\sim 6$ from the HUDF. Their results showed a factor of 6 decline
in star formation rate from $z\sim 3$ to 6. Even using the most optimistic
escape faction ($f = 1$), their results imply insufficient photons
to ionize the Universe by $z\sim 6$ from stellar sources.

\begin{figure}
\centerline{\psfig{figure=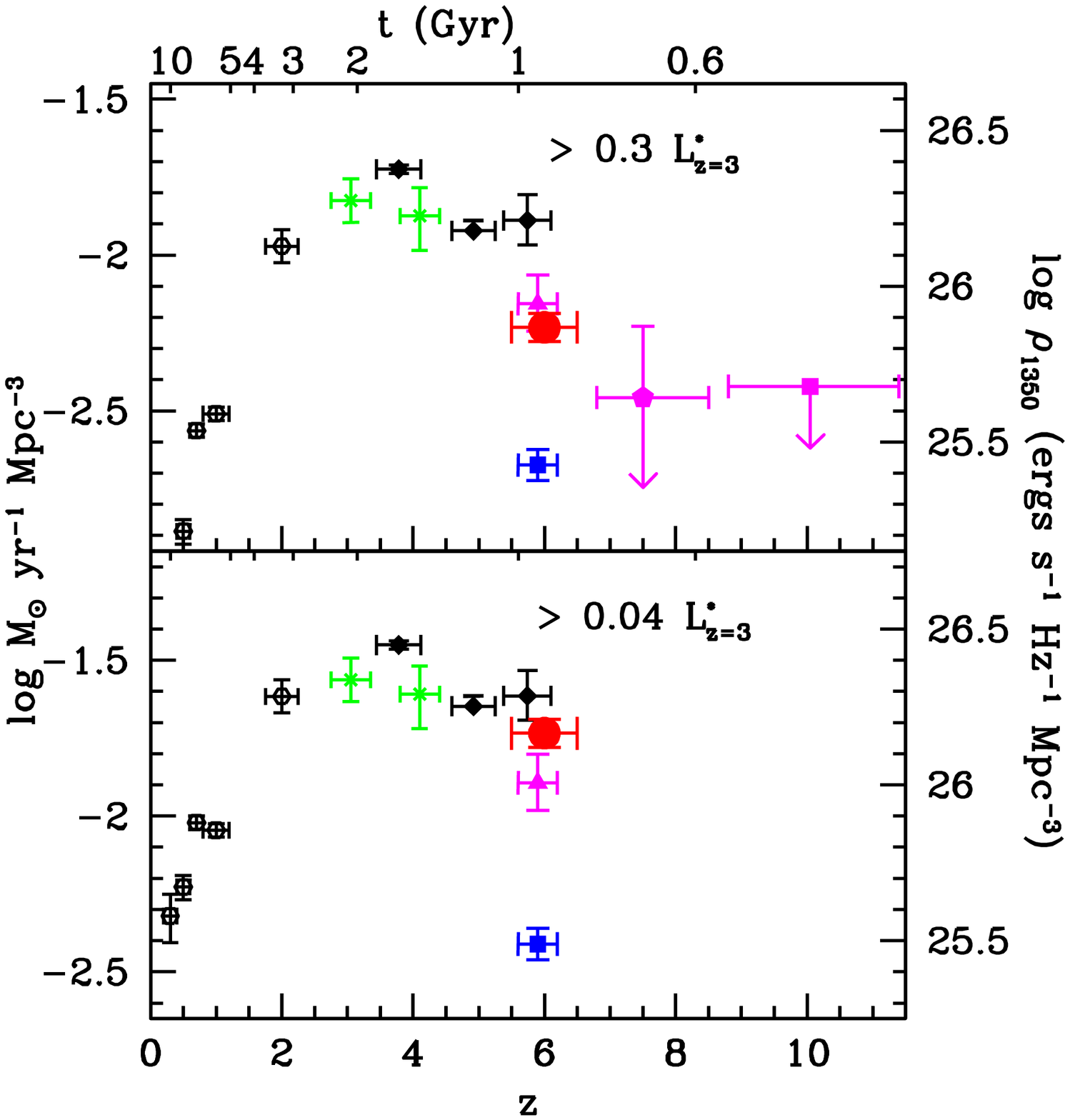,height=30pc}}
\caption{
The cosmic star formation history (uncorrected for
extinction) integrated down to $0.3 L_{z=3} ^{*}$ (top panel) and
$0.04 L_{z=3} ^{*}$ (bottom panel).  These luminosities correspond to
the faint-end limits for $z_{850}$ and $i_{775}$-dropout probes at
$z\sim7-8$ and $z\sim6$, respectively.  The large red circle denotes
determination at $z\sim 6$ from Bouwens et al. (2006).
and compared with
previous determinations by Schiminovich et al.\ (2005) (open squares),
Steidel et al.\ (1999) (green crosses), Giavalisco et al.\ (2004b)
(black diamonds), Bouwens et al.\ 2004a (magenta triangle), Bunker et al. (2004) (blue
square), Bouwens et al.\ (2004b) (magenta pentagon), and Bouwens et
al.\ (2005) (magenta square).
The figure is
divided into two panels to illustrate how much stronger the evolution
is at the bright end of the LF ($\gtrsim 0.3 L_{z=3}^{*}$) than it is
when integrated to the $i$-dropout faint-end limit
($0.04L_{z=3}^{*}$).
Adapted from Bouwens et al. (2005).
}
\label{bouwens}
\end{figure}

Bouwens et al. (2005) combined all the available datasets that include
506 $i$-dropout ($z\sim 6$) galaxy candidates to construct a
luminosity function.  Figure \ref{bouwens} summarizes the
determination of the evolution of star-formation rate at both the
luminous and faint end. The star formation rate begins to decline
modestly at $z>5$.  The shape of the luminosity function has also
evolved significantly from $z\sim 3$ to 6.  However, the integrated UV
luminosity (down to 0.04 $L^*_{z=3}$) did not evolve significantly.
Their results support the previous claim that dwarf galaxies provide
the majority of ionizing photons at $z\sim 6$, sufficient to
ionize the Universe.

Most recently, Kashlinsky et al. (2005) have reported the detection of
a significant excess in the Spitzer IRAC bands above that expected
based on known galaxy counts. They interpret this excess as being due
to population III stars forming at $z > 10$.  These stars would
clearly contribute to early reionization, although the magnitude of
the contribution remains uncertain due to the difficulty of the
measurement, and the uncertainty of the origin of this mid-infrared
background.

\subsection{Dusty and big galaxies at high redshift}

Recent observations have revealed two surprising types of galaxies at
$z > 6$. First, millimeter observations of the host galaxies of the
highest redshift QSOs reveal large dust masses ($\sim 10^8$ M$_\odot$)
in $\sim 30\%$ of the sources (Bertoldi et al. 2003a).  And in one
case, J1148+5251 at $z =6.42$, millimeter line emission from CO has
been detected, indicating a large molecular gas mass ($\sim 2\times
10^{10}$ M$_\odot$; Walter et al. 2003, 2004; Bertoldi et al. 2003b).
Second, Eyles et al. (2005) and Mobasher et al. (2005) present Spitzer
observations of $z\sim 6 - 6.5$ galaxies.  These objects show SEDs
characteristic of post-starburst galaxies, with stellar masses between
$10^{10} - 10^{11} M_{\odot}$ and stellar ages of several hundred
Myrs.  Reddening is also infered by Chary, Stern, \& Eisenhardt (2005)
in one galaxy, implying a dusty galaxy ($\rm A_v \sim 1$).
These results require active star formation ($> 10^2$ M$_\odot$
year$^{-1}$) starting at $z\sim 8 - 15$. These objects are capable of
ionizing large volumes at high-redshift, and could have played an
important role in the early reionization of the IGM, depending on
their currently ill-constained space density (Panagia et al. 2005;
Barkana \& Loeb 2005c).

\subsection{X-ray photons and decaying sterile neutrios}

High-energy X-ray photons from mini-quasars or supernova remnants at
high-redshift could also contribute significantly to the reionization
process (Oh 2001; Venkatesan et al. 2001; Madau et al. 2004; Ricotti
\& Ostriker 2004). These photons have long mean-free-paths, and could
pre-ionize the IGM significantly, providing the optical depth required
by WMAP polarization measurement, and produce a rapid ending of
reionization at $z\sim 6$ as suggested by the Gunn-Peterson
measurements.  However, Dijkstra et al. (2004) and Salvaterra et al. (2005) show that the hard
X-rays from these same sources would produce a present-day soft X-ray
background.  They calculate the X-ray background if such photons
dominated the reionization budget, and conclude that models with
accreting black holes (a combination of luminous and faint AGN) would
overproduce the observed X-ray background by a large factor. A
population dominated by mini-quasars could still partially ionize the
IGM at $z>6$, but its contribution could be severely constrained if
the X-ray background is further resolved into discrete sources.

Finally, Hansen \& Haiman (2004) proposed reionization by decaying
sterile neutrinos at high-redshift. However, Mapelli \& Ferrara
(2005) showed that by satisfying limits on the X-ray, optical and
near-IR cosmic background, the total Thomson optical depth from
sterile neutrinos is small, and they could have only played a minor
role in reionization.  Chen \& Kaminkowski (2003) and Pierpaoli (2003)
show that the number of decays needed to reionize the universe at
$z\sim 20$ will produce a CMB inconsistent with observations.

Furlanetto et al. (2004b) show that low frequency observations of the
power spectrum of brightness temperature fluctuations in the HI 21cm
line from the neutral IGM during reionization will be able to
differentiate between uniform versus HII region dominated
reionization.

\section{HI 21cm probes of Reionization}

The 21cm line of neutral hydrogren presents a unique probe of the
evolution of the neutral intergalactic medium (IGM), and cosmic
reionization.  Furlanetto \& Briggs (2004) point out some of
the advantages of using the HI line in this regard: (i) unlike
Ly$\alpha$ (ie. the Gunn-Peterson effect), the 21cm line does not
saturate, and the IGM remains `translucent' at large neutral
fractions (Carilli et al. 2002). And (ii) unlike CMB polarization studies,
the HI line provides full three dimensional (3D) information on
the evolution of cosmic structure, and the technique involves
imaging the neutral IGM directly, and hence can easily distiguish
between different reionization models (Furlanetto et al. 2004b). HI 21cm
observations can be used to study the evolution of cosmic
structure from the linear regime at high redshift (ie.
density-only evolution), through the non-linear, `messy
astrophysics' regime associated with luminous source formation. As
such, HI measurements are sensitive to structures ranging from
very large scales down to the source scale set by the cosmological
Jeans mass, thereby making 21cm the ``richest of all cosmological
data sets" (Barkana \& Loeb 2005b).

Many programs have been initiated to study the HI 21cm signal from
cosmic reionization, and beyond.  The largest near-term efforts are
the Mileura Wide Field Array
(MWA\footnote{web.haystack.mit.edu/arrays/MWA/LFD/index.html}), the
Primeval Structure Telescope (PAST\footnote{web.phys.cmu.edu/past/}),
and the Low Frequency Array (LOFAR\footnote{www.lofar.org/}).  These
telescopes are being optimized to study the power spectrum of the HI
21cm fluctuations. The VLA-VHF\footnote{cfa-www.harvard.edu/dawn/}
system is designed specifically to set limits on the cosmic \strom
spheres around $z\sim 6$ to 6.4 SDSS QSOs. In the long term the Square
Kilometer Array (SKA\footnote{www.skatelescope.org/}) should have the
sensitivity to perform true three dimensional imaging of the neutral
IGM in the 21cm line during reionization.  And at the lowest
frequencies ($< 80$ MHz), the Long Wavelength Array
(LWA\footnote{lwa.unm.edu/index.shtml}), and eventually the Lunar
array (LUDAR; Corbin et al 2005; Maccone 2005), are being designed for
the higher $z$ HI 21cm signal from the PGM.

In this section we summarize the current theories and capabilities for
detecting the neutral IGM during, and prior to, reionization using the
HI 21cm line (see also Carilli 2005; Furlanetto, Oh, \& Briggs 2006).

\subsection{The physics of the neutral IGM}

The physics of radiative transfer of the HI 21cm line
through the neutral IGM have been considered in detail by many authors
(Scott \& Rees 1990; Bharadwaj \& Ali 2005; Furlanetto \& Briggs 2004;
Zaldariagga et al. 2004b; Santos et al. 2005; Tozzi et al. 1999;
Madau et al. 1997; Morales 2005).
We review only the basic results here.

In analogy to the Gunn-Peterson effect for Ly$\alpha$ absorption
by the neutral IGM, the optical depth, $\tau$, of the neutral
hydrogen to 21cm absorption for our adopted values of
the cosmological parameters is:
\begin{equation}
\tau
 = {{3c^3 \hbar A_{10} n_{HI}}\over{16 k_B \nu_{21}^2 T_S H(z)}} \\
  \sim 0.0074 \frac{x_{HI}}{T_S} (1+\delta)(1+z)^{3/2}
  [H(z)/(\frac{dv}{dr})],
  \label{tauofz}
\end{equation}
\noindent where $A$ is the Einstein coefficient and
$\nu_{21}$ = 1420.40575 MHz (eg. Santos et al. 2005).
This equation shows immediately the rich physics involved in
studying the HI 21cm line during reionization, with $\tau$ depending
on the evolution of cosmic over-densities, $\delta$ (predominantly in
the linear regime), the neutral fraction, $x_{HI}$
(ie. reionization),  the HI excitation, or spin, temperature,
$T_S$, and the velocity structure, $\frac{dv}{dr}$, including
the Hubble flow and peculiar velocities.

In the Raleigh-Jeans limit, the observed
brightness temperature (relative to the CMB)
due to the HI 21cm line at a frequency
$\nu = \nu_{21}/(1+z)$, is given by:
\begin{equation}
T_B  \approx ~  \frac{T_S - T_{\rm CMB}}{1+z} \, \tau
\label{eq:dtb} \\
 \approx ~  7 (1+\delta) x_{HI} (1 - \frac{T_{CMB}}{T_S})
(1+z)^{1/2} ~ \rm{mK},
\end{equation}
\noindent The conversion factor from brightness temperature to
specific intensity, I$_\nu$, is given by:
$I_\nu = \frac{2 k_B}{(\lambda_{21} (1+z))^2} T_B =
22 (1+z)^{-2} T_B$ Jy deg$^{-2}$.
Equation 6 shows that for
$T_S \sim T_{CMB}$ one expects no 21cm signal.
When $T_S >> T_{CMB}$, the brightness temperature
becomes independent of spin temperature. When
$T_S << T_{CMB}$, we expect a strong negative
(ie. absorption) signal against the CMB.

Tozzi et al. (1999) point out that the HI excitation temperature will
equilibrate with the CMB on a timescale $\sim
\frac{3\times10^5}{(1+z)}$ year, in absence of other effects.
However, collisions, and resonant scattering of Ly$\alpha$ photons, can
drive $T_S$ to the gas kinetic temperature, $T_K$ (Wouthuysen 1952;
Field 1959; Hirata 2005).  Zygelman (2005) calculates that, for the mean IGM
density, collisional coupling between $T_S$ and $T_K$ becomes
significant for $z \ge 30$.  Madau et al. (1997) show that resonant
scattering of \lya photons will couple $T_S$ and $T_K$ when $J_\alpha
> 9\times 10^{-23} (1+z)$ erg cm$^{-2}$ s$^{-1}$ Hz$^{-1}$ sr$^{-1}$,
or about one Ly$\alpha$ photon per every two baryons at $z=8$.

The interplay between the CMB temperature, the kinetic temperature,
and the spin temperature, coupled with radiative transfer, lead to a
number of interesting physical regimes for the HI 21cm signal (Ali
2005; Barkana \& Loeb 2004; Carilli 2005): (I) At $z> 200$ equilibrium
between $T_{CMB}$, $T_K$, and T$_{S}$ is maintained by Thomson
scattering off residual free electrons and gas collisions.  In this
case $T_S = T_{CMB}$ and there is no 21cm signal.  (II) At $z \sim 30$
to 200, the gas cools adiabatically, with temperature falling as
$(1+z)^2$, ie. faster than the (1+z) for the CMB. However, the mean
density is still high enough to couple $T_S$ and $T_K$, and the HI
21cm signal would be seen in absorption against the CMB (Sethi 2005).
(III) At $z \sim 20$ to 30, collisions can no longer couple $T_K$ to
$T_S$, and $T_S$ again approaches $T_{CMB}$. However, the \lya photons
from the first luminous objects (Pop III stars or mini-quasars), may
induce local coupling of $T_K$ and $T_S$, thereby leading to some 21cm
absorption regions (Cen 2006).
At the same time, Xrays from these same objects
could lead to local IGM warming above $T_{CMB}$ (Chen et al. 2003).
Hence one might expect a patch-work of regions with no signal,
absorption, and perhaps emission, in the 21cm line.  (IV) At $z \sim
6$ to 20 all the physical processes come to play. The IGM is being
warmed by hard Xrays from the first galaxies and black holes (Loeb \&
Zaldarriaga 2004; Barkana \& Loeb 2004; Ciardi \& Madau 2003), as well
as by weak shocks associated with structure formation (Furlanetto,
Zaldarriaga, \& Hernquist 2005; Wang \& Hu 2005), such that $T_K$ is
likely larger than $T_{CMB}$ globally (Furlanetto et al. 2004b).
Likewise, these objects are reionizing the universe, leading to a
fundamental topological change in the IGM, from the linear evolution
of large scale structure , to a bubble dominated era of HII regions
(Furlanetto et al. 2005a).

\subsection{Capabilities to detect HI 21cm signals from reionization}

The HI 21cm signature of the neutral IGM during, and prior to,
reionization can be predicted analytically, using a standard
Press-Schechter type of analysis of linear structure formation, plus
some recipes to approximate non-linear evolution (Santos et al. 2005;
Zaldariagga et al. 2004a; Ali 2005; Bharadwaj \& Ali 2005; Wang \& Hu
2005; Gnedin \& Shaver 2004), or through the use of numerical
simulations (Ciardi \& Madau 2003; Fulanetto, et al. 2004b; Wang et
al. 2005).

\begin{figure}
\centerline{\psfig{file=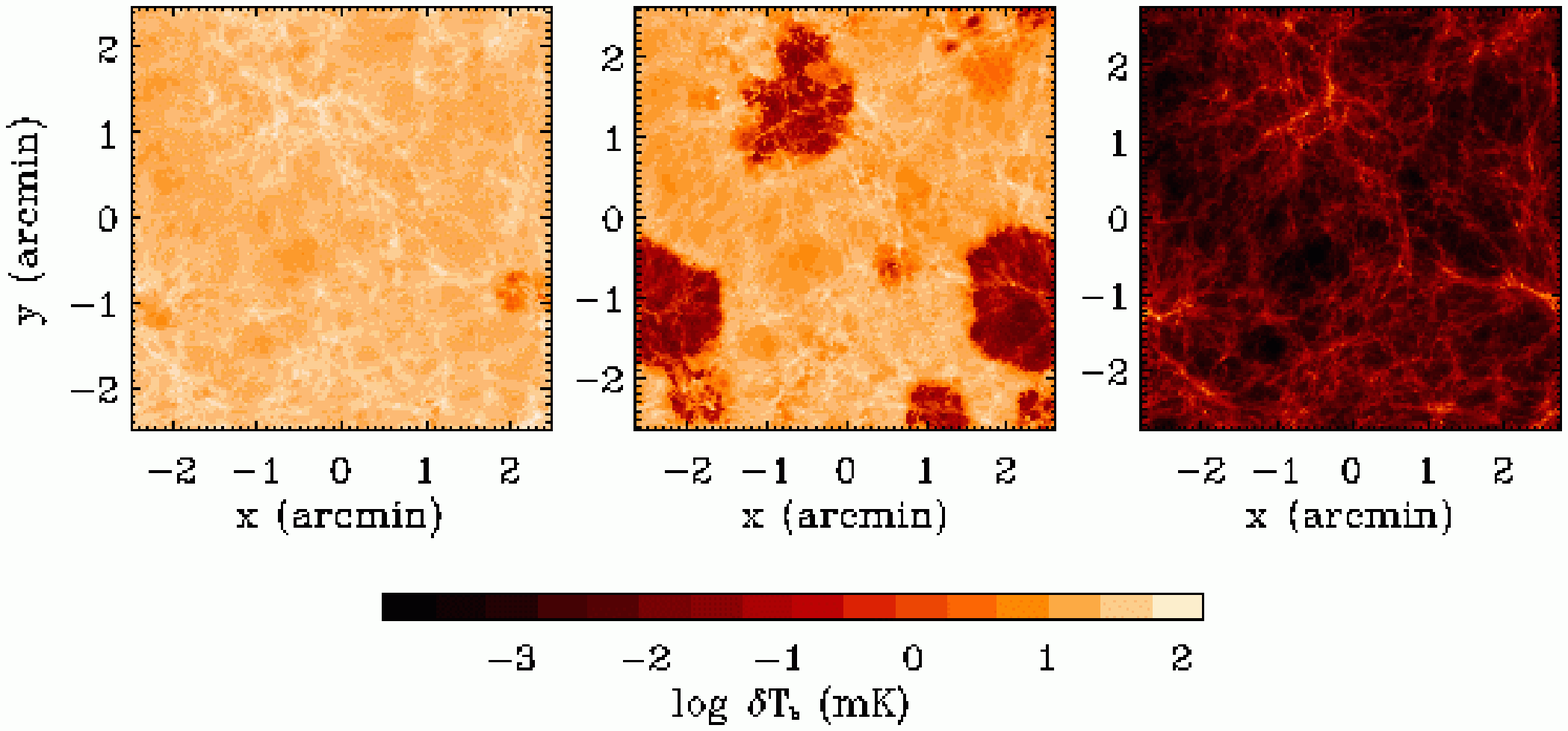,width=6in}}
\caption{The simulated HI 21cm brightness temperature distribution
during reionization at $z = 12$, 9, 7 (Furlanetto et al. 2004a;
Zaldariagga et al. 2004a).}
\label{HIsim}
\end{figure}

{\bf Tomography:} Figure \ref{HIsim} shows the expected evolution of
the HI 21cm signal during reionization based on numerical simulations
(Zaldaraiagga et al. 2004a; Furlanetto et al. 2004a; Furlanetto et
al. 2004b).  They find that the mean HI signal is about $T_B \sim 25$
mK prior to reionization, with fluctuations of a few mK on arcmin
scales due to linear density evolution.  In this simulation, the HII
regions caused by galaxy formation are seen in the redshift range $z
\sim 8$ to 10, reaching scales up to 2$'$ (frequency widths $\sim 0.3$
MHz $\sim 0.5$ Mpc physical size). These regions have (negative)
brightness temperatures up to 20 mK relative to the mean HI signal.
This corresponds to 5$\mu$Jy beam$^{-1}$ in a 2$'$ beam at 140 MHz.

The point source rms sensitivity (dual polarization)
in an image from a synthesis radio telescope  is given by:
\begin{equation}
{\rm rms} = ~
(\frac{1.9}{(\Delta\nu_{kHz} t_{hr})^{0.5}})
(\frac{T_{sys}}{\epsilon_{eff}A_{ant}N_{ant}}) ~~ \rm{Jy ~ beam^{-1}}
\end{equation}
\noindent where $\Delta \nu$ is the channel width in kHz, $t$ is the
integration time in hours,
$A_{ant}$ is the collecting area of each element in the array (m$^2$),
$\epsilon_{eff}$ is
the aperture efficiency, and $N$ is the number of elements ($\epsilon
A N$ = total effective collecting area of the array). At low frequency,
the system temperature, $T_{sys}$, is dominated by the synchrotron
foreground, and behaves as: $T_{FG} \sim 100
(\frac{\nu}{200 \rm{MHz}})^{-2.8}$, in the coldest regions of the sky.
Consider the SKA, with an effective collecting area of one square
kilometer at 140 MHz, distributed over 4 km, $T_{sys} = 300$ K,
a channel width of 0.3 MHz, and integrating for one month. The
rms sensitivity is then 1.3$\mu$Jy beam$^{-1}$, with a beam FWHM $\sim
2'$. This square kilometer array will be adequate to perform
three dimensional imaging of the average structure of the IGM during
reionization.

\begin{figure}
\psfig{file=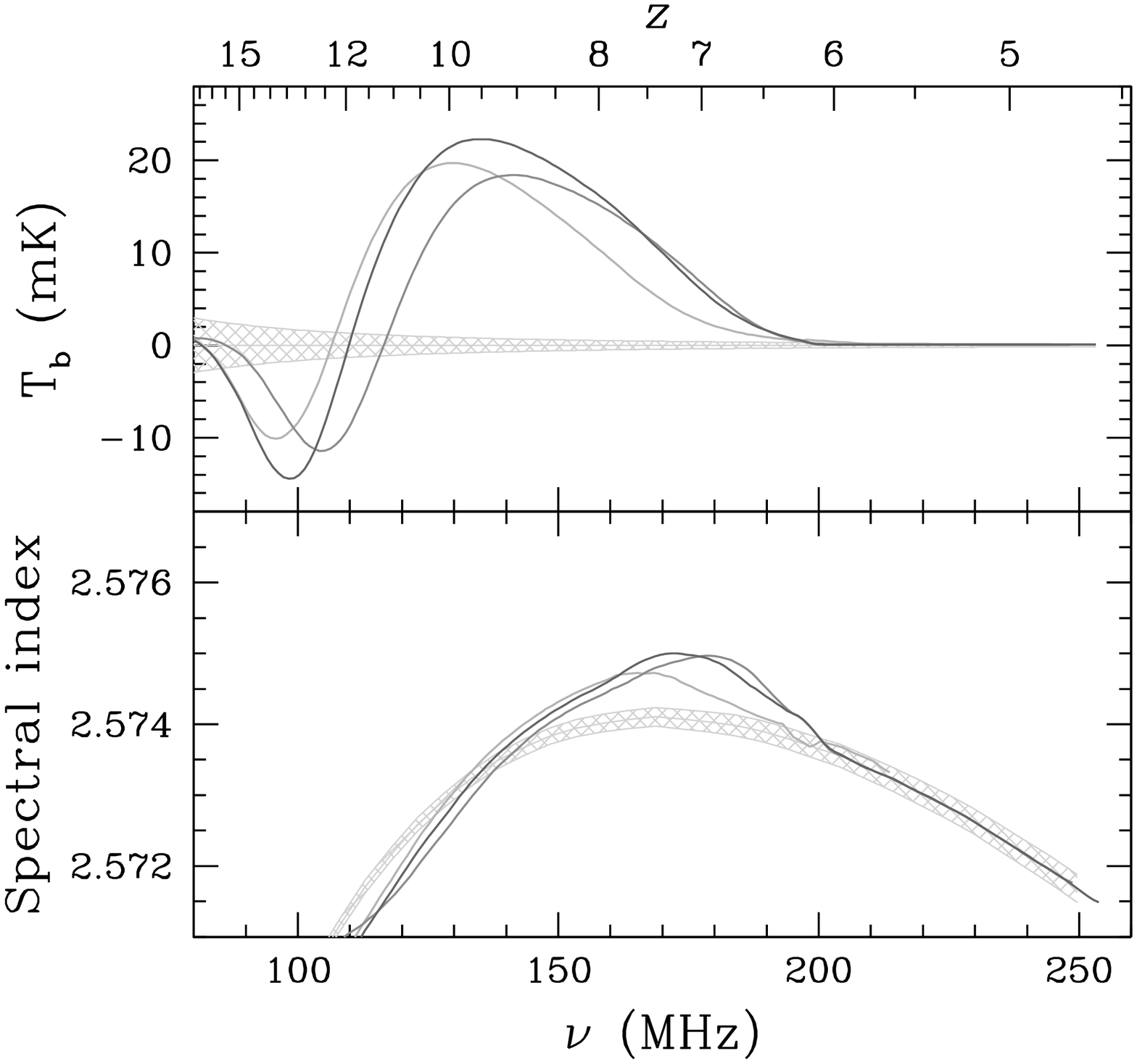,width=2.0in}
\vskip -1.8in
\hspace*{2.1in}
\psfig{file=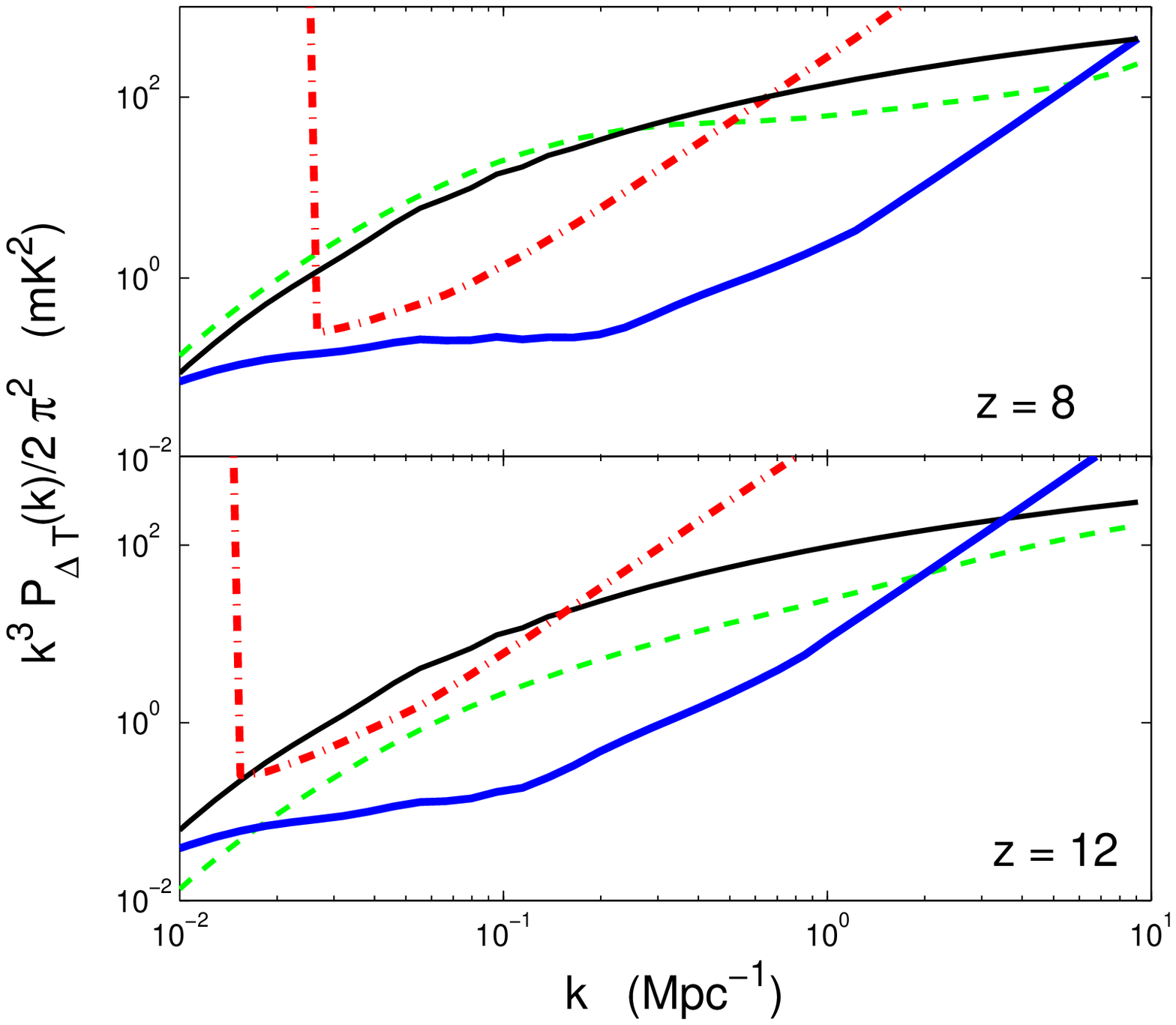,width=2.0in}
\vskip -1.82in
\hspace*{4.2in}
\psfig{file=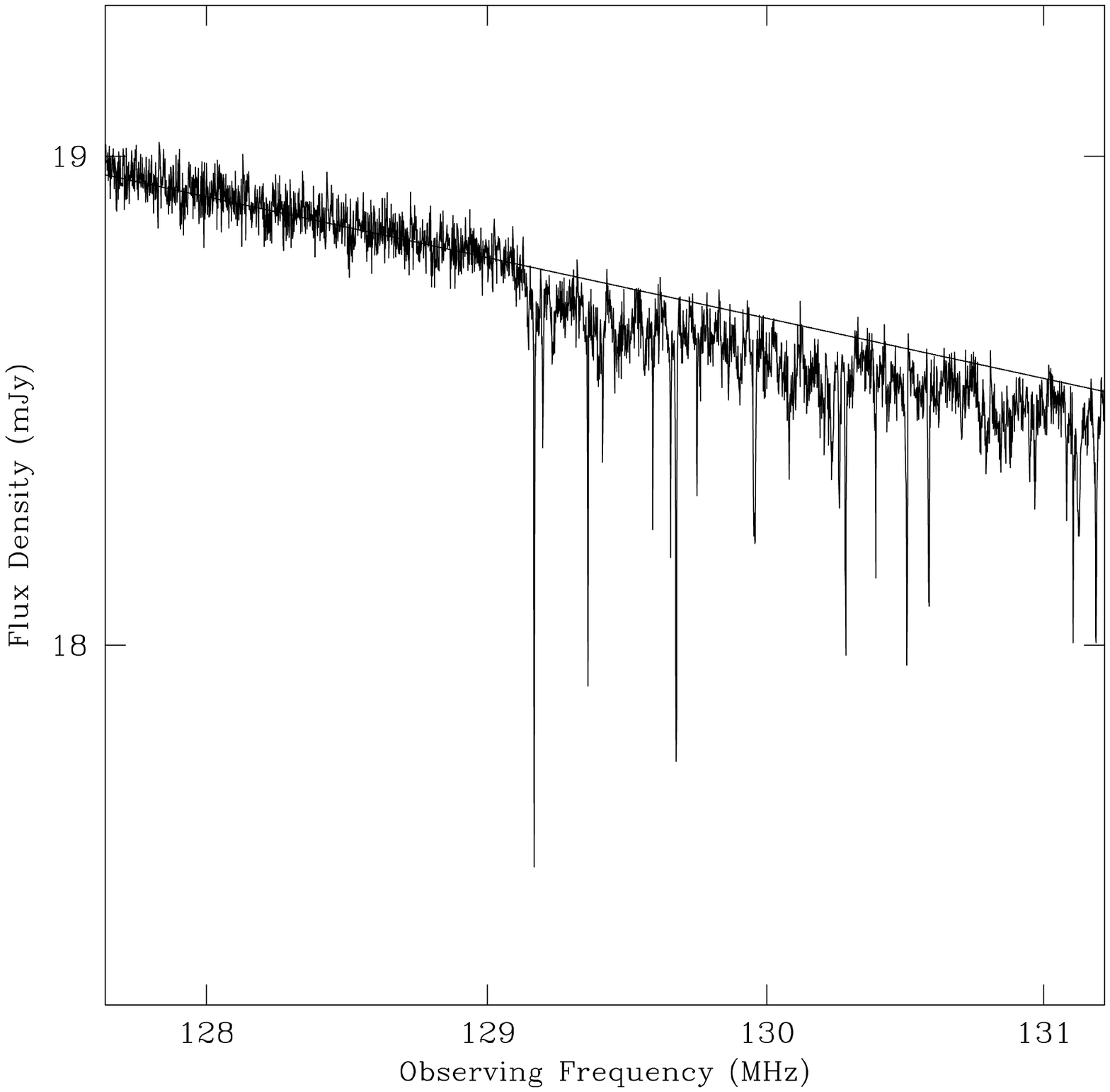,width=1.8in}
\hskip 0.2in
\caption{{\bf Left}: Global (all sky) HI signal from reionization
(Gnedin \& Shaver 2003). The shaded region shows the expected thermal
noise in a carefully controlled experiment.
{\bf Center:} Predicted HI 21cm brightness temperature power spectrum
 (in log bins) at redshifts 8 and 12 (McQuinn et al. 2006).  The thin
black line shows the signal when density fluctuations dominate.  The
dashed green line shows the predicted signal for $\bar{x}_i = 0.2$
at $z=12$, and $\bar{x}_i = 0.6$ at $z=8$,
in the Furlanetto et al. (2004) semi-analytic model. The thick blue line
shows the SKA sensitivity in 1000hrs. The thick red dot-dash show the
sensitivity of the pathfinder experiment LOFAR. The cutoff at low k is
set by the primary beam.
{\bf Right}:  The simulated SKA spectrum of a radio continuum source
at $z = 10$ (Carilli et al. 2002). The straight line is the intrinsic
power law (synchrotron) spectrum of the source. The noise curve
shows the effect of the 21cm line in the neutral IGM, including
noise expected for the SKA in a 100 hour integration.}
\label{HIsignal}
\end{figure}

Unfortunately, the nearer term low frequency path finder arrays will
have $\le 10\%$ the collecting of the SKA, and will likely not be able
to perform such direct 3D imaging (see Table 1 in Carilli 2005).
However, these near term experiments should have enough sensitivity to
probe the neutral IGM in other ways.  Figure \ref{HIsignal} shows
three possible HI 21cm line signatures that might be observed prior to
the SKA.

{\bf Global signal:} The left panel in Figure \ref{HIsignal}
shows the latest predictions of
the global (all sky) increase in the background temperature due to the
HI 21cm line from the neutral IGM (Gnedin \& Shaver 2003).  The
predicted HI emission signal peaks at roughly 20 mK above the
foreground at $z \sim 10$. At higher redshift, prior to IGM warming,
but allowing for Ly$\alpha$ emission from the first luminous objects,
the HI is seen in absorption against the CMB.  Since this is an all
sky signal, the sensitivity of the experiment is independent of
telescope collecting area, and the experiment can be done using small
area telescopes at low frequency, with well controlled frequency
response (Subrahmanyan, Chippendale, Ekers 2005). Note that the line
signal is only $\sim 10^{-4}$ that of the mean foreground continuum
emission at $\sim 150$ MHz.

{\bf Power spectra:} The middle panel in Figure \ref{HIsignal} shows
the predicted power spectrum of spatial fluctuations in the sky
brightness temperature due to the HI 21cm line (Mcquinn et al. 2006).
For power spectral analyses the sensitivity is greatly enhanced
relative to direct imaging due to the fact that the universe is
isotropic, and hence one can average the measurements in annuli in the
Fourier (uv) domain, ie. the statistics of fluctuations along an
annulus in the uv-plane are equivalent (Santos et al. 2005;
Zaldariagga et al. 2004a; Bharadwaj \& Ali 2005; Morales 2005;
Subrahmanyan et al. 1998; Bowman et al. 2005).  Moreover, unlike the
CMB, HI line studies provide spatial and redshift information, and
hence the power spectral analysis can be performed in three
dimensions.

Figure \ref{HIsignal} includes the power spectrum arising from linear
density fluctuations following the dark matter, plus the effect of
reionization.  The signature of reionization can be seen as a bump in
the power spectrum above the density-only curve due to the formation
of HII regions.  The rms fluctuations at $z = 10$ peak at about 10 mK
rms on scales $\ell \sim 5000.$\footnote{Or $\theta \sim
\frac{180^o}{\ell} = 2.2'$, or $\rm k\sim \frac{\ell}{10^4} Mpc^{-1} =
0.5 Mpc^{-1}$ comoving.} A factor 2 to 4 increase in the rms might be
expected over this naive (Poisson) calculation due to gas peculiar
velocities, ie.  infall into superclusters and along filaments
(Barkana \& Loeb 2005a, Ali 2005; but cf. Kuhlen et al. 2005), and
possible clustering of luminous sources, ie. biased galaxy formation
(Furlanetto et al.  2005a; Santos et al. 2005).

Also included in Figure \ref{HIsignal} are the noise power spectra for
near and longer term radio arrays. The noise power spectrum, in
mK$^2$, in standard spherical harmonic units, $\frac{\ell(\ell +1)}{2
\pi} C_\ell^N$, for an array with (roughly) uniform coverage of the uv
plane, is given by:
\begin{equation}
C_\ell^N = \frac{T_{sys}^2 (2\pi)^3}{\Delta \nu t f_c^2 \ell_{max}^2}
\end{equation}
\noindent where $\Delta \nu$ is the channel width in Hz, $t$ is
the integration time in seconds, $f_c$ = areal covering factor of the
array $= N_{ant} \frac{A_{ant}}{A_{tot}}$, $A_{tot}$ is the total area
of the array defined by the longest baseline
($A_{tot} \def \pi (\frac{b_{max}}{2})^2$),
and $\ell_{max}$ is set by
the longest baseline ($b_{max}$) in the array:
$\ell_{max}= 2\pi b_{max}/\lambda$ (Santos et al. 2005; for a three dimensional
generalization, see Morales 2005).
For example, the SKA will have 30$\%$ of its
collecting area inside 1 km, or $f_c = 0.3$ out to 1 km.
Typical noise values for near-term arrays are $\sim 1$ to 10
mK rms in the range $\ell = 10^3$ to $10^4$, for long integrations
(see Figure \ref{HIsignal}). This sensitivity
is adequate to constrain the power
spectrum of the HI 21cm fluctuations, even if true images of the
emission cannot be constructed.

A point of debate has been the fraction of HI in collapsed objects, as
opposed to the diffuse IGM (Iliev et al. 2003). This fraction has a complex
dependence on structure formation history.  Oh \& Mack (2003) and
Gnedin (2004) conclude that the majority of the HI
during reionization will remain
in the diffuse phase, while Ahn et al. (2005) argue that most of the HI
will reside in mini-halos. Iliev et al (2003), and Ahn (2005)
have considered the HI 21cm power spectrum due to clustering of such
minihalos at high redshift.  For a beam size of 1$'$, and channel
width of 0.2 MHz, they predict 3$\sigma$ brightness temperature
fluctuations due to clustered minihalos $\sim 7$mK at $z = 8.5$,
decreasing to 2mK at $z=20$.

{\bf Absorption toward discrete radio sources:} A interesting
alternative to emission studies is the possibility of studying smaller
scale structure in the neutral IGM by looking for HI 21cm absorption
toward the first radio-loud objects (AGN, star forming galaxies, GRBs)
(Carilli et al. 2004b). The rightpanel of Figure \ref{HIsignal}
shows the predicted HI 21cm absorption signal toward a high redshift
radio source due to the `cosmic web' prior to reionization, based on
numerical simulations (Carilli, Gnedin, Owen 2002).  For a source at
$z = 10$, these simulations predict an average optical depth due to
21cm absorption of about 1$\%$, corresponding to the `radio
Gunn-Peterson effect', and about five narrow (few km/s) absorption
lines per MHz with optical depths of a few to 10$\%$. These latter
lines are equivalent to the Ly $\alpha$ forest seen after
reionization, and correspond to over-densities evolving in the linear
regime ($\delta \le 10$).  Fulanetto \& Loeb (2002), and Oh \& Mack
(2003) predict a similar HI 21cm absorption line density due to gas in
minihalos as that expected for the 21cm forest.

Fundamental to absorption studies is the existence of radio
loud sources during the EoR.  This question has been considered in
detail by Carilli et al. (2002), Haiman et al. (2004), and
Jarvis \& Rawlings (2005).  They show that current
models of radio-loud AGN evolution predict between 0.05 and 1 radio
sources per square degree at $z > 6$ with $\rm S_{150MHz} \ge 6$ mJy,
adequate for EoR HI 21cm absorption studies with the SKA.

{\bf Cosmic \strom spheres:} While direct detection of the typical
structure of HI and HII regions may be out of reach of the near-term
EoR 21cm telescopes, there is a chance that even this first generation
of telescopes will be able to detect the rare, very large scale HII
regions associated with luminous quasars near the end of reionization
(\S 4).  The expected signal is $\sim 20$mK $\times x_{HI}$ on a scale
$\sim 10'$ to 15$'$, with a line width of $\sim 1$ to 2 MHz (Wyithe \&
Loeb 2004b). This corresponds to 0.5 $\times x_{HI}$ mJy beam$^{-1}$,
for a 15$'$ beam at $z \sim 6$ to 7.  Kohler et al. (2005) calculate
the expected spectral dips due to large HII regions around luminous
quasars ($> 2\times 10^{10}$ L$_\odot$) during the EoR.  In a typical
spectrum from 100 to 180 MHz with a 10$'$ beam they predict on average
one relatively deep (-2 to -4 mK) dip per LoS on this scale.  Wyithe
et al. (2005b) perform a similar calculation using the evolution of
the bright QSO luminosity function to predict the number of HII
regions around active QSOs at $z > 6$. They conclude that there should
be roughly one SDSS-type HII region around an active QSO (physical
radius $>$ 4Mpc) per 400 deg$^2$ field per 16 Mhz bandwidth at $z \sim
6$, and one $R \ge 2$ Mpc region at $z \sim 8$, and up to $\sim
100\times$ more fossil HII regions due to non-active AGN,
depending on the duty cycle.

{\bf Beyond reionization:} Recent calculations of the brightness
temperature fluctuations due to the 21cm line have extended to
redshifts higher than reionization, $z > 20$.  Barkana \& Loeb (2004;
2005a) predict the power spectrum of fluctuations during the era of
`\lya coupling' ($z \sim 20$ to 30; see also Kuhlen et al. 2005 and
Cen 2006).  Brightness temperature fluctuations can be due to emission
from clustered minihalos, plus enhanced by absorption against the CMB
by the diffuse IGM, with a predicted rms $\sim$ 10mK for $\ell \sim
10^5$, due to a combination of linear density fluctuations, plus
Poisson (`shot') noise and biasing in the Ly$\alpha$ source
(ie. galaxy) distribution.

Loeb \& Zaldarriaga (2004) go even further in redshift, to $z > 50$ to
200. In this regime the HI generally follows linear density
fluctuations, and hence the experiments are as clean as CMB studies,
and $T_K < T_{CMB}$, so a relatively strong absorption signal might be
expected.  They also point out that Silk damping, or photon diffusion,
erases structures on scales $\ell > 2000$ in the CMB at recombination,
corresponding to comoving scales = 22 Mpc. The HI 21cm measurements
can explore this physical regime at $z \sim 50$ to 300. The predicted
rms fluctuations are 1 to 10 mK on scales $\ell= 10^3$ to 10$^6$ (0.2$^o$
to 1$''$).  These observations could provide the best tests of
non-Gaussianity of density fluctuations, and constrain the running
power law index of mass fluctutions to large $\ell$, providing important
tests of inflationary structure formation. Sethi (2005) also suggests
that a large global signal, up to -0.05 K, might be expected for this
redshift range.

Unfortunately, the sky background is very hot $> 10^4$ K at these low
frequencies ($< 50$ MHz), and the predicted signals are orders of
magnitude below the expected sensitivities of even the biggest planned
low frequency arrays in the coming decade.

\subsection{Observational challenges}

{\bf Foregrounds:} The HI 21cm signal from reionization must be
detected on top of a much larger synchrotron signal from foreground
emission (\S 9.2). This foreground includes discrete radio galaxies,
and large scale emission from our own Galaxy, with relative
contributions of about 10\% and 90\%, respectively.  The expected HI
21cm signal is about $10^{-4}$ of the foreground emission at 140 MHz.

di Matteo et al. (2002) show that, even if point sources can be
removed to the level of 1 $\mu$Jy, the rms fluctutions on spatial
scales $\le 10'$ ($\ell \ge 1000$) due to residual radio point sources
will be $\ge 10$ mK just due to Poisson noise, increasing by a factor
100 if the sources are strongly clustered (see also di Matteo, Ciardi,
\& Miniati 2004; Oh \& Mack 2003).

A key point is that the foreground
emission should be smooth in frequency, predominantly the sum of
power-law, or perhaps gently curving, non-thermal spectra.
A number of complimentary approaches have been presented for
foreground removal (Morales, Bowman, \& Hewitt 2005).  Gnedin \&
Shaver (2003) and Wang et al (2005) consider fitting smooth
spectral models (power laws or low order polynomials in log space) to
the observed visibilities or images. Morales \& Hewitt (2003) and
Morales (2004) present a 3D Fourier analysis of the measured
visibilities, where the third dimension is frequency.  The different
symmetries in this 3D space for the signal arising from the noise-like
HI emission, versus the smooth (in frequency) foreground emission, can
be a powerful means of differentiating between foreground emission and
the EoR line signal.  Santos et al. (2005), Bharadwaj \& Ali (2005),
Zaldariagga et al. (2004a) perform a similar analysis, only in the
complementary Fourier space, meaning cross correlation of spectral
channels. They show that the 21cm signal will effectively decorrelate
for channel separations $> 1$ MHz, while the foregrounds do not.  The
overall conclusion of these methods is that spectral decomposition
should be adequate to separate synchrotron foregrounds from the HI
21cm signal from reionization at the mK level.

{\bf Ionosphere:} A second potential challenge to low frequency
imaging over wide fields is phase fluctuations caused by the
ionosphere.  These fluctuations are due to index of refraction
fluctuations in the ionized plasma, and behave as $\Delta \phi \propto
\nu^{-2}$.  Morever, the typical `isoplanatic patch', or angle over
which a single phase error applies, is a few to 10 degrees (physical
scales of 10's km in the ionosphere), depending on frequency (Cotton
et al. 2004; Lane et al. 2004).  Fields larger than the isoplanatic
patch will have multiple phase errors across the field, and hence
cannot be corrected through standard (ie. single solution) phase
self-calibration techniques.  New wide field self-calibration
techniques, involving multiple phase solutions over the field, or a
`rubber screen' phase model (Cotton et al. 2004; Hopkins et al. 2003),
are being developed that should allow for self-calibration over wide
fields.

{\bf Interference:}
Perhaps the most difficult problem facing low frequency radio
astronomy is terrestrial (man-made) interference (RFI). The relevant
frequency range corresponds to 7 to 200 MHz ($z= 200$ to 6). These are
not protected frequency bands, and commercial allocations include
everything from broadcast radio and television, to fixed and mobile
communications.

Many groups are pursuing methods for RFI mitigation and excision (see
Ellingson 2004). These include: (i) using a reference horn, or one
beam of a phased array, for constant monitoring of known, strong, RFI
signals, (ii) conversely, arranging interferometric phases to produce
a null at the position of the RFI source, and (iii) real-time RFI
excision using advanced filtering techniques in time and frequency, of
digitized signals both pre- and post-correlation. The latter requires
very high dynamic range (many bit sampling), and very high frequency
and time resolution.

In the end, the most effective means of reducing interference is to go
to the remotest sites.  The MWA and PAST have selected sites in remote
regions of Western Australia, and China, respectively, because of
known low RFI environments. Of course, the ultimate location would be
the far-side of the moon.

The technical challenges to HI 21cm observations of reionization are
many. Use of spectral decomposition to remove the foregrounds requires
careful control of the synthesize beam as a function of frequency,
with the optimal solution being a telescope design where the
synthesized beam is invariant as a function of frequency (Subrahmanyan
et al. 2005).  High dynamic range front ends are required to avoid
saturation in cases of strong interference, while fine spectral
sampling is required to avoid Gibbs ringing in the spectral response.
The polarization response must be stable and well calibrated to remove
polarized foregrounds (Haverkorn et al. 2004). Calibation in the
presence of a structured ionospheric phase screen requires new wide
field calibration techniques.  The very high data rate expected for
many element ($\ge 10^3$) arrays requires new methods for data
transmission, cross correlation, and storage.  At the
lowest frequencies, $\le 20$MHz or so, where we hope to study the
PGM, phase fluctuations and the opacity of the
ionosphere becomes problematic, leading to the proposed LUDAR project
on the far side of the moon. The far side of the moon is also the best
location in order to completely avoid terrestrial interference.

\section{Summary}

\begin{figure}
\centerline{\psfig{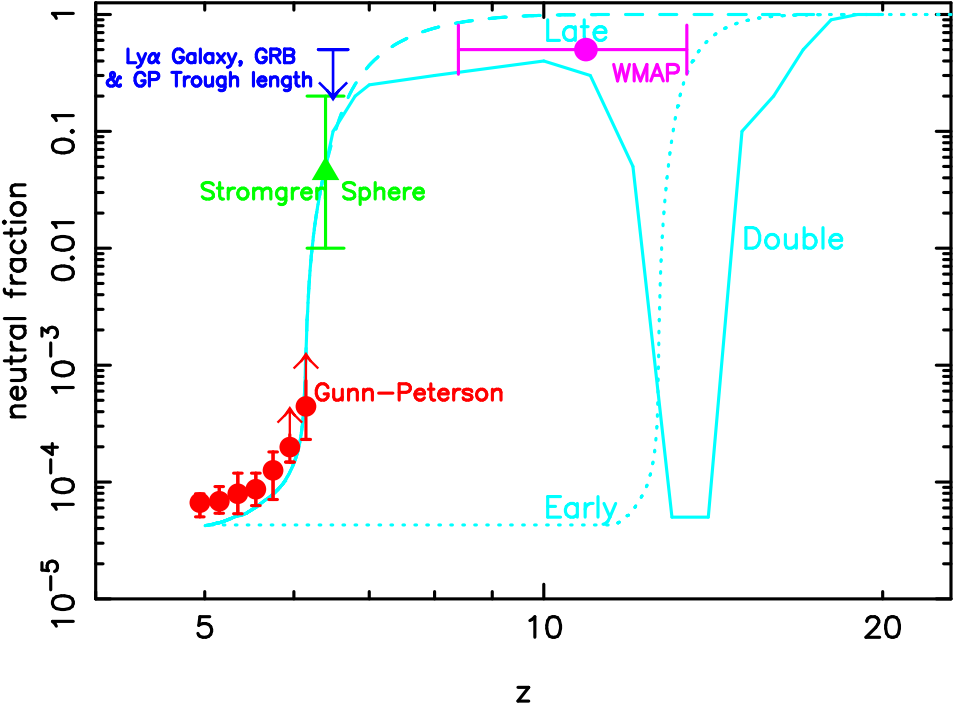}}
\caption{The volume averaged neutral fraction of the IGM
versus redshift using various techniques. The dash line
shows the fiducial model of Gnedin (2004) with late reionization
at $z=6 - 7$, the solid line shows an idealized
model with double reionization as described in Cen (2003a), and
the dotted line illustrates the
model with early reionization at $z\sim 14$.}
\label{FHI}
\end{figure}

The few years since the seminal theoretical reviews of Barkana \& Loeb
have seen dramatic progress in determining observational constraints
on cosmic reionization. Figure \ref{FHI} shows the current limits on
the cosmic neutral fraction versus redshift.  The observations paint
an interesting picture.  On the one hand, studies of GP optical depths
and variations, and the GP `gap' distribution, as well as of the
thermal state of the IGM at high $z$, and of cosmic \strom spheres and
surfaces around the highest redshift QSOs, suggest a qualitative
change in the state of the IGM at $z \sim 6$. These data indicate a
significant neutral fraction, $x_{HI} > 10^{-3}$, and perhaps as high
as 0.1, at $z \ge 6$, as compared to $x_{HI} \le 10^{-4}$ at $z <
5.5$.  The IGM characteristics at this epoch are consistent with the
end of the `percolation' stage of reionization (Gnedin \& Fan 2006).
On the other hand,
transmission spikes in the GP trough and study of the evolution of
\lya galaxy luminosity function indicate a neutral fraction smaller
than 50\% at $z\sim 6.5$.
Moreover, the measurement of the large scale
polarization of the CMB implies a significant ionization fraction
extending to higher redshifts, $z \sim 15$ to 20.
Moreover, the measurement of the
large scale polarization of the CMB suggests a significant
ionization fraction extending to higher redshifts, $z \sim 11\pm3$.

We emphasize that all these measurements have implicit assumptions
and uncertainties, as discussed throughout. Indeed, the GP effect and
CMB large scale polarization studies can be considered complimentary
probes of reionization, with optical depth effects limiting GP studies
to the end of reionization, while CMB studies are weighted toward the
higher redshifts, when the densities were higher.  The data argue
against a simple reionization history in which the IGM remains largely
neutral from $z\sim 1100$ to $z\sim 6-7$, with a single phase
transition at $z\sim 6$ (the ``late'' model in Figure \ref{FHI}), as
well as against a model in which the Universe reaches complete
ionization at $z\sim 15-20$ and remained so ever since (the ``early''
model in Figure \ref{FHI}).  These facts, combined with the large line
of sight variations at the end of reionization as indicated by
Gunn-Peterson measurements, suggest a more extended reionization
history. Interestingly, the latest theoretical models with
reionization caused by Population II star formation are consistent
with both GP optical depth and WMAP CMB polarization measurement
(e.g. Gnedin \& Fan 2006).  Current data do not present strong
evidence for a dominant contribution by metal-free Population III star
formation at $z>15$ to reionization (Haiman \& Bryan 2006), and
have weak constraints on, although do not require, models with
multiple episodes of reionization (the ``double'' model in Figure
\ref{FHI}), which were suggested by a possible high CMB Thomspon
optical depth based on early WMAP measurements.

Overall, we agree with the assessment of Furlanetto et al. (2004b)
that reionization is less an event than a process, extended in both
time and space.  Much emphasis has been given to the determination of
the `reionization redshift'.  However, current observations suggest
that, unlike recombination, which occurs over a very narrow range in
fractional redshift, $z_{recomb} = 1089 \pm 1$, cosmic reionization
occurs over a fairly large fractional redshift, from $z \sim 6$ up to
as high as $z \sim 14$. The redshift of HII region overlap might be
quite different than the redshift at which the IGM is largely
($>50\%$) ionized.  It may also differ in different parts of the
universe.  The implication is that the formation of the first liminous
objects (stars and AGN) occurs in a `twilight zone', heavily obscured
at (observed) optical wavelengths by a partially neutral IGM, and
observable only at near-IR through radio wavelengths, and in the hard
Xrays.

We also consider the sources responsible for reionization.  Current
data are consistent with star forming galaxies, in particular,
relatively low luminosity galaxies, as being the dominant sources of
reionizing photons.  However, again, there are a number of important,
and poorly constrained, parameters in these calculations, including
the IMF and the UV escape fraction, and we certainly cannot rule-out a
significant contribution from mini-QSOs.

Lastly, we show that low frequency radio telescopes currently under
construction should be able to make the first direct measurements of
HI 21cm emission from (and/or absorption by) the neutral IGM.  These
observations will present a clear picture of the reionization process,
and the evolution of the neutral IGM into the dark ages.

\vskip 0.2in

Acknowledgements: CC thanks the Max-Planck-Gesellschaft and the
Humboldt-Stiftung for support through the Max-Planck-Forschungspreis,
K. Menten and the MPIfR for their hospitality, and S. Furlanetto
for extensive comments.  XF thanks supports from NSF grant AST
03-07384, a Sloan Research Fellowship and Packard Fellowship for
Science and Engineering.  BK thanks Nathan Miller for his
contributions to this review and Steven Furlanetto, Manoj Kaplinghat,
and Meir Shimon for helpful comments on this manuscript. CMBFAST was
used in the preparation of this review.
The National Radio Astronomy Observatory (NRAO) is
a facility of the National Science Foundation, operated under
cooperative agreement by Associated Universities, Inc.

\end{document}